\documentclass[twocolumn,aps,prx,showpacs,amsmath,superscriptaddress,longbibliography,notitlepage]{revtex4-2}
\usepackage{xcolor}
\usepackage{amssymb}
\usepackage{capt-of}
\usepackage{graphicx}
\usepackage{dcolumn}
\usepackage{float}
\usepackage{bm}
\usepackage{amsmath}
\usepackage{lipsum}
\usepackage{enumitem}
\usepackage{mathtools}
\usepackage{comment}
\usepackage{tikz}
\usepackage{soul}
\usepackage{nicematrix}
\usepackage{amsmath, braket}
\usepackage[normalem]{ulem}

\usepackage[
    colorlinks=true,
    linkcolor=red,
    citecolor=blue,
    urlcolor=blue
]{hyperref}

\newcommand{\be}{\begin{equation}}
\newcommand{\ee}{\end{equation}}

\DeclareMathOperator{\Tr}{Tr}

\newcommand{\beq}{\begin{equation}}
\newcommand{\eeq}{\end{equation}}
\newcommand{\beqn}{\begin{eqnarray}}
\newcommand{\eeqn}{\end{eqnarray}}

\usepackage[noframe]{showframe}
\usepackage{framed}

\renewenvironment{shaded}{%
  \MakeFramed{\advance\hsize-\width \FrameRestore\FrameRestore}}%
 {\endMakeFramed}
\definecolor{shadecolor}{gray}{0.75}

\newcommand{\leo}[1]{{\color{blue} \footnotesize (\textsf{Leo}) \textsf{\textsl{#1}}}}

\usepackage{tikz}

\tikzset{
    coltensor/.style={
        draw,
        rounded corners,
        minimum width=1.0cm,
        minimum height=.60cm,
        inner sep=1pt
    },
    tleg/.style={
        line width=.45pt
    }
}

\begin{document}

\title{Lieb-Schultz-Mattis Constraints for Quantum Channels: A Spacetime-Duality View}

\author{Sarang Gopalakrishnan}
\affiliation{Department of Electrical and Computer Engineering,
Princeton University, Princeton, NJ 08544, USA}

\author{Yu-Jie Liu}
\affiliation{Center for Theoretical Physics - a Leinweber Institute, Massachusetts Institute of Technology, Cambridge, MA 02139, USA}

\author{Tsung-Cheng Lu}
\affiliation{Joint Center for Quantum Information and Computer Science,
University of Maryland, College Park, Maryland 20742, USA}
\affiliation{Department of Physics and Astronomy, Center for Quantum Research and Technology,
University of Oklahoma, Norman, OK 73069, USA}

\author{Frank Pollmann}

\affiliation{Technical University of Munich, TUM School of Natural Sciences, Physics Department, 85748 Garching, Germany}

\author{Yizhi You}
\affiliation{Department of Physics, Northeastern University, Boston, MA, 02115, USA}
\thanks{Authors are listed in alphabetical order.}

\date{\today}

\begin{abstract}
Quantum anomalies strongly constrain the possible behavior of many-body systems. A prime example is the Lieb-Schultz-Mattis (LSM) theorem, which relates UV symmetry and filling constraints to IR features of the energy spectrum and ground-state structure. Here, we ask how LSM constraints shape dynamical signatures and temporal correlations in open quantum systems. Using a spacetime duality, we show that the Liouvillian of a $d$-dimensional repeated quantum channel with a mixed anomaly between strong $S$ and weak $G$ symmetry can be mapped to a $(d{+}1)$-dimensional mixed-state symmetry-protected topological (mSPT) phase. Under this correspondence, the initial and steady states of the channel are identified with boundary states of the higher-dimensional mSPT in the presence of bulk projection. We further introduce the twisted Renyi-$N$ correlator (TRNC) as a probe of temporal correlations in the channel and demonstrate that it is dual to the mSPT strange correlator, providing a direct bulk-boundary route to diagnose long-range temporal order implied by the LSM anomaly. Finally, we identify the \textit{Liouvillian singular spectrum}, rather than the Liouvillian spectrum itself, as a more fundamental diagnostic of quantum anomalies, and show that it is dual to the operator entanglement spectrum of the mSPT. 
\end{abstract}

\maketitle

\tableofcontents

\section{Introduction}

Quantum anomalies are powerful tools for constraining quantum many-body systems, providing insight into their low-energy properties from the underlying ultraviolet (UV) structure. The Lieb-Schultz-Mattis (LSM) theorem exemplifies this phenomenon, revealing universal low-energy behavior dictated by UV symmetries \cite{lieb1961two,affleck1986proof,oshikawa2000commensurability}. Regardless of the microscopic Hamiltonian, the combined spatial symmetry (G), internal symmetry (S), and filling conditions (e.g., spin per unit cell) impose nontrivial constraints on the infrared (IR) properties of many-body spin systems, enforcing a degenerate or gapless energy spectrum. Recent studies have clarified that the LSM theorem is deeply tied to quantum anomalies. As elucidated in Refs.~\cite{lu2024lieb,jian2018lieb,else2014classifying,cho2017anomaly,po2017lattice}, under renormalization group flow, the spatial symmetry \( G \) maps to an effective internal symmetry \( \tilde{G} \) in a coarse-grained, enlarged unit cell. The resulting effective theory can be viewed as a quantum field theory with a 't Hooft anomaly under \( \tilde{G} \times S \) symmetry, which again prevents a unique gapped ground state. This unified framework connects the LSM theorem to SPT boundary physics and informs the search for quantum spin liquids and deconfined quantum critical points.

The LSM theorem has recently been extended to open quantum systems~\cite{kawabata2024lieb,zhou2025reviving,you2024intrinsic,zang2024detecting,wang2024anomaly,sun2024holographic,hsin2024anomalies} driven by decoherence and dissipation~\cite{bao2023mixed,fan2023diagnostics,lee2023quantum,zhang2022strange,lessa2025mixed,huang2025interaction,ding2024boundary,hsin2024anomalies,sala2024spontaneous,li2023intrinsically,wang2024anomaly,li2026mixed,panahi2026quantum}. 
Ref.~\cite{kawabata2024lieb} formulates an LSM theorem for open quantum systems driven by Lindbladian evolution: for a mixed state \(\rho\) of a spin-\(1/2\) chain with strong \(U(1)\) (or \(D_2\)) symmetry and weak lattice-translation symmetry, the Liouvillian spectrum must be either gapless or degenerate. A subtle point, however, is that in open quantum systems, a gapless or degenerate Liouvillian spectrum can arise rather generically. In particular, strongly \(G\)-symmetric quantum channels always have degenerate steady states associated with different fixed \(G\)-charge sectors~\cite{gu2024spontaneous}, while strongly \(U(1)\)-symmetric channels always exhibit a gapless Liouvillian spectrum due to hydrodynamic modes~\cite{huang2024hydro}. Thus, any channel with strong symmetry should exhibit a degenerate Liouvillian spectrum even in the absence of an anomaly; likewise, a channel with strong continuous symmetry should display a gapless Liouvillian spectrum even without an anomaly. The Liouvillian spectrum alone is therefore not a sharp fingerprint of the anomaly, since its gapless or degenerate features can already be enforced by the CPTP condition together with strong symmetry. This makes it essential to identify what additional salient features the LSM constraint imposes on open systems beyond the Liouvillian spectrum itself.

In closed systems, the LSM theorem supports several generic features beyond the spectral gap: it triggers (quasi) long-range order in both space and temporal directions, and it can be measured and detected by inserting a twisted boundary condition \cite{else2014classifying,kapustin2014anomalies}. Whether these features admit a direct generalization in open quantum systems remains widely unexplored \cite{lee2023quantum,ma2024topological,sarma2025effective,sala2025entanglement,wang2024anomaly,wang2023intrinsic,xu2025average,you2024intrinsic,zang2024detecting,huang2026anomaly,yi2026order,liu2026local,divi2026local,zang2026topological,yi2026order}.
In this context, a central question is whether the anomaly underlying the LSM constraint enforces additional structures on the long-time dynamics generated by Lindblad evolution or repeated quantum channels. More precisely, when an anomalous quantum channel is applied repeatedly, what dynamical consequences follow? In particular, does the anomaly enforce distinctive temporal correlations \cite{gu2024spontaneous,guo2024new,guo2025strong,moharramipour2024symmetry,salo_steady_state_strong_sym_2025,liu2025parent,lu2025holographic}?

In this work, we study open-system dynamics under the Lieb-Schultz-Mattis (LSM) constraint by analyzing the stroboscopic time evolution generated by repeated applications of a quantum channel. Using the spacetime duality of Ref.~\cite{lu2025spacetime}, we reinterpret the Liouvillian operator, equivalently the quantum-channel transfer matrix \(Q\) in superoperator space, as a spatial building block of a higher-dimensional wavefunction.
Within this correspondence, any \(d\)-dimensional quantum channel subject to an LSM constraint, characterized by a mixed anomaly between strong \(S\) and weak \(G\) symmetries, is mapped to a \((d{+}1)\)-dimensional mixed-state symmetry-protected topological (\(m\)SPT) phase with the same strong \(S\) and weak \(G\) symmetries~\cite{ma2022average,lee2025symmetry,sarma2025effective,zhang2022strange, ma2024topological,guo2025strong,ando2024gauge,zhang2025probing}. Under this duality, temporal correlations of the quantum channel, which relate input and output states under repeated evolution, are mapped to boundary correlations of the \((d{+}1)\)-dimensional \(m\)SPT in the presence of bulk projection. 
This connection follows from the correspondence between the path integral of an anomalous theory in $d$ dimensions and the wavefunction of an SPT state in $(d+1)$ dimensions, whose spatial boundary realizes the same anomalous symmetry action. This perspective was introduced in Refs.~\cite{you2014wave,xu2013wave,bi2015classification} to characterize anomalies and SPT phases in thermal equilibrium. Here, we extend this framework to nonequilibrium settings in open systems.




To make the correspondence precise, we will introduce a temporal correlation function for open quantum systems: the twisted Rényi-\(N\) correlator (TRNC) of a quantum channel~\cite{sala2025entanglement}. We show that the TRNC under spacetime duality is mapped to the strange correlator of the higher-dimensional \(m\)SPT. Consequently, long-range order in the strange correlator implies long-range temporal order in the TRNC. In parallel, we show that the Liouvillian singular spectrum, defined as the singular-value spectrum of the quantum-channel transfer matrix \(Q\), is dual to the operator entanglement spectrum of the higher-dimensional \(m\)SPT. It follows that an LSM constraint requires the \textit{singular spectrum of \(Q\)} to be gapless or degenerate. This distinction is important because gaplessness or degeneracy in the spectrum of \(Q\) itself can arise from entirely conventional mechanisms in strongly symmetric channels, such as the completely positive and trace-preserving (CPTP) condition or hydrodynamic modes. By contrast, the singular spectrum, equivalently the eigenvalue spectrum of \(Q^\dagger Q\), is less explored but provides a sharper and more faithful diagnostic of the underlying anomaly. 

The rest of the manuscript is organized as follows.In Sec.~\ref{sec:main_lsm}, we review the basic properties of quantum channels and the LSM constraints on their steady states. We then elucidate the spacetime-dual picture, which maps a 1d quantum channel subject to an LSM constraint onto a 2d mSPT state. In Sec.~\ref{sec:Sing}, we explore the Liouvillian singular spectrum and demonstrate that it cannot have a unique dominant eigenvector under an LSM constraint. In Sec.~\ref{sec:general}, we discuss the general formulation of LSM constraints in quantum channels. 

\section{LSM constraints for quantum channels via spacetime duality}\label{sec:main_lsm}

\subsection{Basics for quantum channels and LSM}\label{sec:basic}

To set the stage, we analyze the infinite-time evolution of an open quantum system governed by Markovian dynamics. By discretizing time, the continuous Lindblad evolution over each interval can be described as a local quantum channel $\mathcal{E}$:
\begin{align}
\label{decohere_1general}
     \mathcal{E}(\rho_0) = \sum_i K_i\, \rho_0\, K_i^{\dagger},
\end{align}
where the set of Kraus operators $\{K_i\}$ satisfies the normalization condition $\sum_i K_i^{\dagger} K_i = \mathbb{I}$.
By iterating this channel, the system may approach a steady state in the long-time limit
\begin{align}
\rho_s = \lim_{N \to \infty} \mathcal{E}^N(\rho_0).
\end{align}

The density matrix and the action of the repeated quantum channel can be conveniently represented using the Choi--Jamiolkowski isomorphism, where a diagonalized density matrix 
\[
\rho = \sum_\nu \lambda_\nu |\nu \rangle \langle \nu |
\]
is mapped to its doubled-state representation
\[
|\rho\rangle\rangle \propto \sum_\nu \lambda_\nu\, |\nu \rangle_1 \otimes | \nu^* \rangle_2,
\]
where $ |\nu \rangle_1 \in \mathcal{H}_1$, and $| \nu^* \rangle_2 \in \mathcal{H}_2$ is defined by taking a complex conjugate of the corresponding $ |\nu \rangle_1$. In other words, a density matrix acting on a physical Hilbert space becomes a state vector $|\rho\rangle\rangle$ (up to normalization) living in the doubled Hilbert space $\mathcal{H}_1 \otimes  \mathcal{H}_2$ as a bilayer system, with $i=1,2$ labeling the layer index. In particular, this representation allows us to treat the quantum channel as a transfer matrix acting on the bilayer system: 
\begin{align}\label{eq:qtransfer}
|\mathcal{E}(\rho)\rangle\rangle \;=\; Q\, |\rho\rangle\rangle,~~~
Q=\sum_i K_i\otimes K^{*}_i
\end{align}
where $|\rho\rangle\rangle$ and $|\mathcal{E}(\rho)\rangle\rangle$ denote the doubled states associated with the input and output density matrices, and $Q$ is the transfer matrix associated with the channel evolution defined in Eq.~\ref{decohere_1general}.

We make a few remarks on the properties of $Q$ and the symmetry action in the doubled Hilbert space: \\

\noindent \textbf{Spectral radius.}  The eigenvalues of the transfer matrix $Q$ are referred to as the \textit{ Liouvillian spectrum}. Since the quantum channel is a completely positive, trace-preserving (CPTP) map, the Liouvillian spectrum ${\lambda_a}$ lies within the unit disk, i.e., $|\lambda_a|\leq 1$. \\

\noindent \textbf{Steady state.} The evolution of a density matrix under the repeated application of a quantum channel is generated by successive powers of the transfer matrix $Q^N$, which approaches a steady state as $N \to \infty$:
\[
|\rho_{s}\rangle\rangle= \lim_{N\to \infty} |\mathcal{E}^N[\rho]~\rangle\rangle =\lim_{N\to \infty}Q^N\, |\rho\rangle\rangle.
\]

In particular, $|\rho_{s}\rangle\rangle$ is the dominant right eigenvector of $Q$, with an eigenvalue one. \\

\noindent \textbf{Left eigenvector.} The left eigenvector with the largest eigenvalue is the identity operator. In the doubled-state representation, this is the symmetric EPR pair
        \[
          \lvert I\rangle\rangle \;=\; \sum_{\alpha} \lvert\alpha\rangle_{1}\,\lvert\alpha^*\rangle_{2}.
        \]

\noindent \textbf{Symmetries of states.} A symmetry can act on a density matrix $\rho$ in two distinct ways. $\rho$ has the \emph{strong} symmetry $S$ generated by the unitary $U_s$ if it satisfies:
\begin{equation}
U_s \rho = e^{i\theta_s}\rho,~~ \rho U_s^{\dagger}=e^{-i\theta_s}\rho  \quad \text{for } s\in S.
\end{equation}
In the double-state representation, this is equivalent to 
\begin{equation}
(U_s \otimes I)  | \rho \rangle\rangle =  e^{i\theta_s}| \rho \rangle\rangle, ~~( I \otimes   U_s^{*} )| \rho \rangle\rangle =  e^{-i\theta_s}| \rho \rangle\rangle. 
\end{equation}
Namely, the doubled state $ | \rho \rangle\rangle$ has an enlarged symmetry $S \times S$. 

On the other hand,  $\rho$ has the \emph{weak} symmetry $W$ generated by the unitary $U_w$ if it satisfies:
\begin{equation}
U_w \rho U^{\dagger}_w= \rho  \quad \text{for } w\in W. 
\end{equation}
In the double-state representation, this is equivalent to 
\begin{equation}
(U_w \otimes U_w^{*})  | \rho \rangle\rangle  = |  \rho \rangle\rangle.  
\end{equation}

\noindent \textbf{Symmetries of channels.} In addition to defining symmetry of quantum states, we can also consider the notion of strong and weak symmetries for quantum channels~\cite{BucaProsen2012,albert:2014,Lieu2020,liu:2024diss}. A quantum channel with \emph{strong} symmetry $S$ generated by the unitary $U_s$ satisfies:
\begin{equation}
(U_s \otimes I)\, Q\, (U_s \otimes I)^{\dagger}   =  (I\otimes U_s^*)\, Q\, ( I \otimes U_s^* )^{\dagger} = Q
\end{equation}
for any $s \in S$. In terms of the channel defined in the original physical Hilbert space, the channel has the strong $S$ symmetry if all the Kraus operators $K_i$ commute with $U_s$. One immediate consequence is that if an input density matrix has the strong symmetry $U_s\rho= \rho$, then the output state $\mathcal{E}(\rho)$ will have the strong $S$ symmetry as well, i.e. $ U_s \mathcal{E}(\rho)=\mathcal{E}(\rho)$.

On the other hand, a quantum channel with \emph{weak} symmetry $W$ generated by the unitary $U_w$ satisfies
\begin{equation}
(U_w \otimes U_w^*)\, Q\, (U_w \otimes U_w^*)^{\dagger} = Q
\end{equation}
for any $w \in W$. In terms of the channel defined in the original physical Hilbert space, this implies that the overall action of the channel commutes with the map given by conjugation of $U_w$. Namely, $ U_w  \mathcal{E}(\rho)  U_w^{\dagger} =  \mathcal{E}(U_w\rho U_w^{\dagger})$. One immediate consequence is that if an input density matrix has the weak symmetry $U_w\rho U_w^{\dagger}=  \rho$, then the output state $\mathcal{E}[\rho]$ will have the weak $W$ symmetry as well, i.e. $ U_w \mathcal{E}(\rho) U_w^{\dagger}  =\mathcal{E}(\rho)$ for any $\rho$. To summarize, if a channel has a strong (weak) symmetry, then the strong (weak) symmetry in the input state is guaranteed to be preserved under the channel dynamics. 

When the symmetries associated with a quantum channel are anomalous~\cite{lessa2025mixed}, the steady states are guaranteed to exhibit certain non-trivial features or long-range correlations corresponding to various symmetry-breaking patterns; see Appendix \ref{app:symmetry} for a general discussion. On the other hand, little is known about whether the symmetry anomaly can be diagnosed directly from the features of quantum channels, rather than inferred solely from its steady states. In the following, we introduce a concrete diagnostic that extracts such dynamical signatures directly inspired by the framework of spacetime duality.

\subsection{Spin-$\tfrac{1}{2}$ chain under repeated quantum channel}\label{sec:LSMchain}

In this subsection, we will employ a spacetime duality that reinterprets the temporal depth \(N\) of a repeated quantum channel as an additional spatial direction. Under this mapping, the transfer matrix \(Q\) of a \(d\)-dimensional quantum channel is recast as the spatial transfer matrix of a \((d+1)\)-dimensional wavefunction.
Within this duality framework, we show that the time correlator of a repeated 1d channel subject to an LSM constraint is mapped to a spatial strange correlator in the dual 2d mixed-state SPT. Therefore, by analyzing the strange correlator and entanglement properties of the dual 2d \(m\)SPT, we can infer the temporal correlations and spectral properties of the channel Liouvillian, represented by the transfer matrix \(Q\).

We begin by formulating a Lieb-Schultz-Mattis (LSM) theorem for open-system dynamics generated by completely positive trace-preserving (CPTP) processes, including both long-time Lindblad evolution and repeated quantum channels. Since Lindblad evolution can always be recast as repeated application of a quantum channel by discretizing the time evolution, we formulate the discussion throughout in the language of quantum channels. For concreteness, this section focuses on a spin-$1/2$ chain with spin-rotation and translation symmetries, corresponding to the familiar setting underlying the Haldane conjecture. Our arguments, however, readily generalize to broader symmetry settings and higher dimensions, as discussed in Sec.~\ref{sec:general}.

\begin{shaded}\label{prop1}
Proposition I: \textit{Consider a spin-$\tfrac{1}{2}$ chain subject to a quantum channel that preserves a strong $SO(3)$ spin-rotation symmetry and a weak lattice-translation symmetry, the associated transfer matrix $\mathcal{Q}$ develops (quasi) long-range order along the temporal direction, diagnosed by the twisted Rényi-$N$ correlator (TRNC).}
\end{shaded}

The \textit{Twisted Rényi-$N$ Correlator} (TRNC) for the transfer matrix $Q$ measures temporal correlations between operator insertions after evolution by $N$ successive applications of the channel:
\begin{align}\label{eq:autoco}
C(N) = \frac{\Tr\left[ O^{\dagger}(x) Q^{N} O(x) Q^{N} \right]}{\Tr\left[ Q^{2N} \right]}.
\end{align}

Here, both $Q$ and $O(x)$ act on the doubled Hilbert space, and the trace is likewise taken over the same space. $O(x)$ is chosen as a charged operator on the 1d chain in the doubled space that carries charge under either the \textit{strong} $SO(3)$ rotation symmetry (such as a spin magnetization operator) or the \textit{weak} translation symmetry (such as the VBS order parameter). The detailed expression of $O(x)$ will be elucidated in Sec.~\ref{sec:renyi}. 

The TRNC probes the temporal correlation function between charged objects along the time trajectory of the repeated quantum channel. The mixed anomaly between the strong $SO(3)$ symmetry and the weak lattice-translation symmetry forbids the corresponding anomalous theory from having short-ranged temporal correlations. Consequently, the TRNC \textit{cannot decay exponentially} with $N$, but must instead either remain of order $\mathcal{O}(1)$ or decay algebraically as $\frac{1}{N^{\alpha}}$. We provide a detailed proof in Sec.~\ref{sec:prooftrnc}.

To elucidate Proposition I, we formulate a spacetime duality for the quantum channel in four steps. A schematic overview is provided in Table~I.

\begin{enumerate}
    \item \textbf{Elementary building block.}
    In Sec.~\ref{sec;couplewire} and Fig.~\ref{dualmain2}-a, we consider a single step of a quantum channel acting on a spin-$\frac12$ chain with strong $\mathrm{SO}(3)$ symmetry and weak lattice-translation symmetry. Exchanging the spatial direction $y$ and time $t$ maps this channel to a wavefunction on a two-leg ladder. The two legs represent the input and output state of the channel at consecutive time steps. We refer to this ladder wavefunction as the \textit{elementary building block}. 

    \item \textbf{From repeated channels to a 2d mSPT.}
    Duplicating $N$ disconnected copies of the 1d channel along the time direction (Fig.~\ref{dualmain2}-c) maps, under the same spacetime rotation, to a spatial stack of elementary building blocks. The resulting two-dimensional state realizes an mSPT protected by strong $\mathrm{SO}(3)$ symmetry and weak lattice translation. Its mSPT character originates from the mixed anomaly of the original 1d quantum channel, as explained in Sec.~\ref{sec;couplewire}.

    \item \textbf{Temporal correlators as strange correlators.}
    Under the duality, temporal correlation functions of the channel, including the TRNC in Eq.~\ref{eq:autoco}, are mapped to spatial strange correlators of the dual 2d mSPT. The operator dynamics of the channel can therefore be diagnosed through the spatial strange correlator of the associated mSPT. We develop this correspondence in Secs.~\ref{sec:renyi} and \ref{sec:prooftrnc}.

\item \textbf{Louvillian singular spectrum and operator entanglement.}
Under the duality, the \textit{singular spectrum} of the 1d Louvillian (the channel's transfer matrix Q) is related to the operator entanglement spectrum of the dual 2d mSPT. Since the latter is constrained to be gapless or degenerate by its mSPT nature, the LSM constraint imposes a corresponding constraint on the singular spectrum of the Louvillian. We explain this relation and present numerical results in Sec.~\ref{sec:Sing}.
\end{enumerate}

\begin{widetext}

\begin{center}
\renewcommand{\arraystretch}{1.2}

\begin{tabular}{|l|l|}
\hline
\textbf{$d$-dimensional repeated quantum channel}
&
\textbf{$(d+1)$-dimensional mSPT}
\\ \hline

Strong $S$ and weak $G$ symmetries with a mixed anomaly
&
mSPT protected by strong $S$ and weak $G$ symmetries
\\ \hline

Time evolution of the quantum channel
&
Transfer matrix of the mSPT under bulk projection
\\ \hline

Twisted R\'enyi-$N$ correlator in time
&
Strange correlator in space
\\ \hline

Input and output states
&
Spatial boundaries of the mSPT under bulk measurement
\\ \hline

Singular spectrum of $Q$
&
Operator entanglement spectrum of the mSPT
\\ \hline
\end{tabular}

\captionof{table}{Summary of the spacetime duality.}
\label{tab:spacetime-duality}

\end{center}

 \begin{figure}[h]
    \centering
\includegraphics[width=0.6\linewidth]{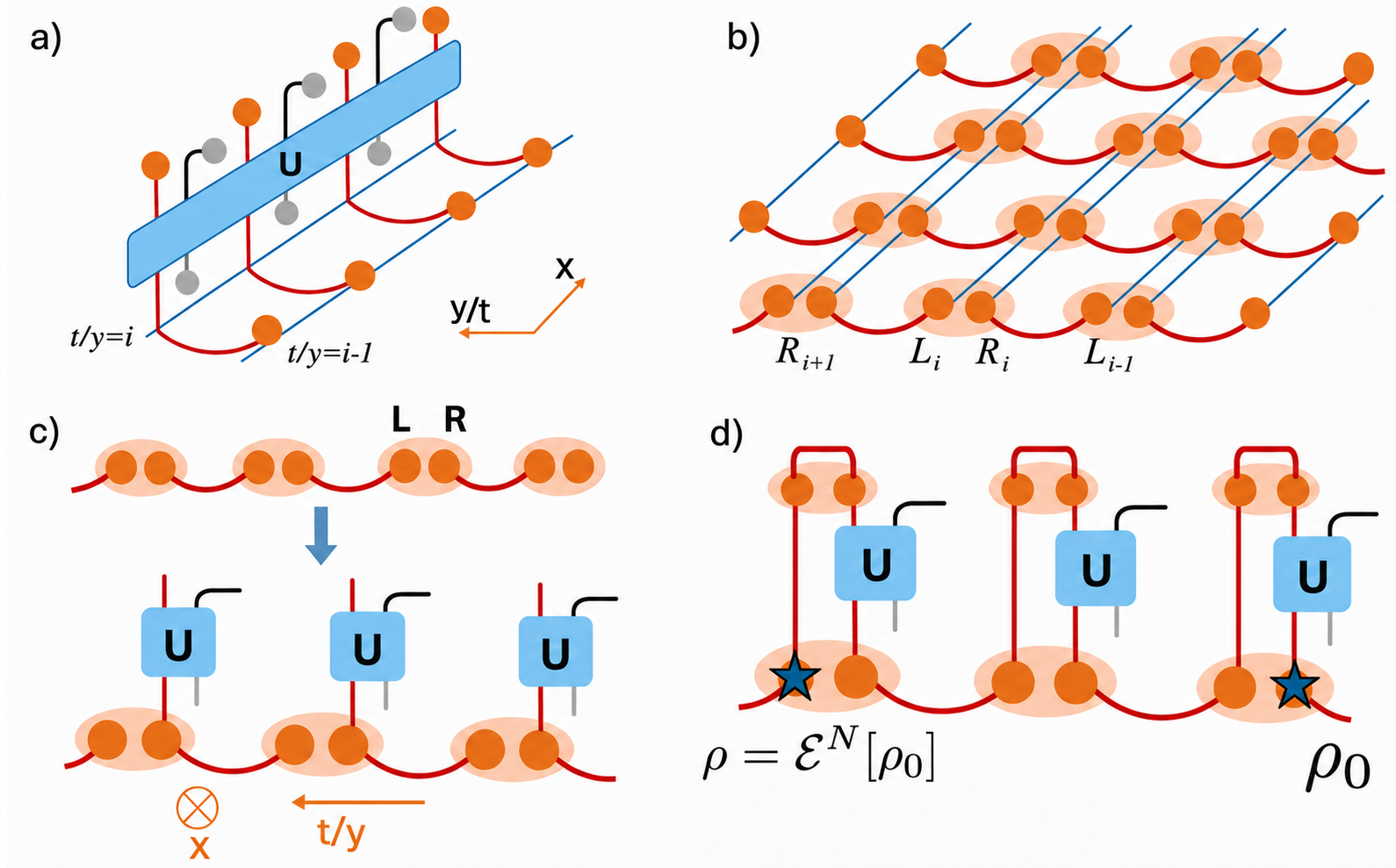}
    \caption{a) Single step of a 1d quantum channel: a system chain (orange) is locally entangled with ancillas (gray) by a unitary circuit \(U\), evolving the state from \(t=i-1\) to \(t=i\). Under the spacetime rotation \(t\leftrightarrow y\), these time slices become adjacent columns, \(y=i-1\) and \(y=i\), of a wavefunction on a ladder.
b)  Coupled-wire construction of the corresponding 2d SPT state in the absence of ancilla. Each column at \(y=i\) contains two spin-\(\frac12\) chains, denoted as \(L_i\) and \(R_i\) components. Coupling the \(R_i\) chain of one column to the \(L_{i-1}\) chain from the neighboring column gaps the bulk and realizes a 2d SPT wavefunction protected by \(\mathrm{SO}(3)\times T_x\).
 c) Construction of the associated 2d mixed-state SPT. The 2d SPT wavefunction is locally coupled to trivial ancilla degrees of freedom (black lines), yielding Eq.~\ref{eq:Phinew}. Tracing out the ancillas then produces the corresponding 2d mSPT.
 d) Projecting \(L_i\) and \(R_i\) in each bulk column onto onsite EPR pairs maps the spatial transfer matrix $Q^N$, after spacetime rotation, to \(N\) successive applications of the channel \(\mathcal E\). The two edges are identified with \(\ket{\rho_0}\!\rangle\) and \(\ket{\mathcal E^N[\rho_0]}\!\rangle\), respectively.}
    \label{dualmain2}
\end{figure}
\end{widetext}

 \subsubsection{From coupled-wire to quantum channels via spacetime duality}\label{sec;couplewire}

To analyze the temporal correlations of a 1d quantum channel, we map a \textit{single update step} to a 1d wavefunction on a ladder through the Choi-Jamiolkowski isomorphism. After a spacetime rotation, the two adjacent time slices \(t=i-1\) and \(t=i\) become neighboring columns \(y=i-1\) and \(y=i\) of a ladder, as shown in Fig.~\ref{dualmain2}-a. The ladder direction \(y\) therefore represents the original time direction.
For the identity channel, the Choi state is
\begin{align}
I
=\sum_{\bm{s}} |\bm{s}\rangle_{t=i}\langle \bm{s}|_{t=i-1}
\xrightarrow{\ \mathrm{Choi}\ }
\sum_{\bm{s}} |\bm{s}\rangle_{y=i}\otimes|\bm{s}\rangle_{y=i-1},
\end{align}
where we choose the local \(S_z\) basis \(s(x)\in\{0,1\}\) and \(\bm{s}=\{s(x)\}_{x=1}^{L}\). Thus, up to normalization, the identity channel produces maximally entangled EPR pairs along the ladder rungs, connecting the input and output qubits at the same spatial coordinate \(x\).

For a general channel, we introduce a Stinespring purification in which the system is coupled to an ancilla chain initialized in the product state
$|\bm{a}\rangle_A
=\otimes_x|0\rangle_A$
through a local unitary \(\hat U^{s,a}\). The ancilla degree of freedom on each site has dimension \(m\); in a minimal Stinespring dilation, \(m\) is the Kraus rank of the local channel. The input ancilla at \(y=i-1\) is fixed in a product state and will be omitted from the following discussion. The resulting dual wavefunction on the ladder is
\begin{align}\label{eq:block}
\ket{\Psi_{\mathrm{block}}}_{i,i-1}
\propto
\sum_{\bm{s}}
\left[
\hat U^{s,a}
\left(
|\bm{s}\rangle_S\otimes|\bm{a}\rangle_A
\right)
\right]_{y=i}
\otimes
|\bm{s}\rangle_{y=i-1}.
\end{align}
We refer to the coupled-ladder wavefunction in Eq.~\ref{eq:block} as an \textit{elementary building block}. In this building block, column \(y=i-1\) contains the input system qubits, whereas column \(y=i\) contains the system and ancilla output of the channel. 
We assume that \(\hat U^{s,a}\) respects \(\mathrm{SO}(3)\) spin rotations and lattice translations \(T_x\), while the ancilla is \(\mathrm{SO}(3)\)-neutral. Tracing out the output ancilla gives the channel in Eq.~\ref{decohere_1general}, with Kraus operators
\begin{align}
K_{\boldsymbol{\alpha}}
={}_A\langle\boldsymbol{\alpha}|\hat U^{s,a}|\bm{a}\rangle_A,
\end{align}
where \(\boldsymbol{\alpha}\) labels a complete output configuration of the ancilla chain. The resulting channel therefore preserves strong \(\mathrm{SO}(3)\) symmetry and weak translation symmetry.

We now stack the \textit{elementary blocks} in Eq.~\ref{eq:block} along the \(y\) direction. This maps a set of \textit{disconnected channel blocks} to a 2d coupled-wire state, whose coordinates are the original spatial direction \(x\) and the channel-time direction \(y\).

We first consider the identity channel and omit the ancilla. Let column \(y=i\) contain two spin-\(\frac12\) chains, denoted by \(L_i\) and \(R_i\). As shown in Fig.~\ref{dualmain2}-b, the identity-channel blocks form EPR pairs between \(R_i\) and \(L_{i-1}\):
\begin{align}\label{eq:Phi}
|\Phi_0\rangle
&=\bigotimes_{i=2}^{N}
|A\rangle_{R_iL_{i-1}},
\\
|A\rangle_{R_iL_{i-1}}
&\propto 
\sum_{\bm{s}}
|\bm{s}\rangle_{R_i}
|\bm{s}\rangle_{L_{i-1}}.
\nonumber
\end{align}
Thus, \(|\Phi_0\rangle\) is a tensor product of maximally entangled building blocks. In each building block, \(L_{i-1}\) and \(R_i\) is dual to the input and output state of the identity channel in a single time step \(t\in \{i-1,i\}\).

The symmetry action on this 2d wavefunction is most transparent in the original Choi basis.
 Let \(U_g\) denote the spin-\(\frac12\) representation of \(g\in\mathrm{SO}(3)\). The output and input chains transform in conjugate representations,
\begin{align}
\mathcal V_g
=
\bigotimes_i
\left(
U_g^{(R_i)}
\otimes
U_g^{*(L_{i-1})}
\right),
\label{eq:microscopic_SO3}
\end{align}
such that
\begin{align}
\left(
U_g^{(R_i)}
\otimes
U_g^{*(L_{i-1})}
\right)
|A\rangle_{R_iL_{i-1}}
=
|A\rangle_{R_iL_{i-1}}.
\end{align}
Translation acts by shifting every chain by one lattice site,
\begin{align}
\mathcal T_x
=
\bigotimes_i
\left(
T_x^{(R_i)}
\otimes
T_x^{(L_{i-1})}
\right).
\label{eq:microscopic_translation}
\end{align}

Eq.~\ref{eq:Phi} is the fixed-point wavefunction of a 2d SPT protected by \(\mathrm{SO}(3)\times T_x\), naturally understood through a coupled-wire construction. Each column \(y=i\) contains two spin-\(\frac12\) wires, \(L_i\) and \(R_i\). Although each individual spin 1/2 chain carries a mixed anomaly between $\mathrm{SO}(3)$ and $T_x$, the entire column (unit cell) \(L_i\oplus R_i\) is anomaly free. The 2d bulk can therefore be symmetrically gapped by pairing \(R_i\) with \(L_{i-1}\) on the neighboring column.
For open boundaries along \(y\), this inter-column pairing leaves one dangling wire, \(R_1\) or \(L_N\), at each boundary. These boundary wires retain the LSM anomaly and cannot be trivially gapped without breaking \(\mathrm{SO}(3)\times T_x\), which consequently realize the anomalous edges of the 2d SPT state. This is the standard coupled-wire construction, in which coupled anomalous 1d wires lead to 2d SPT phase~\cite{jian2018lieb,you2014wave,bi2015classification}. In Appendix~\ref{app:wzwintro}, we elaborate on this argument using the Euclidean path-integral formulation of WZW theory~\cite{xu2013wave,jian2018lieb,metlitski2018intrinsic,yang2018dyonic}.


We next restore the channel ancilla. For each elementary block, the output chain \(R_i\) is coupled to an ancilla chain \(A_i\), initialized in the translation-invariant product state
\begin{align}
|\bm a\rangle_{A_i}
=
\bigotimes_x|0\rangle_{A_i}.
\end{align}
After including the ancilla, the 2d wavefunction is:
\begin{align}
|\Phi\rangle
&=
\bigotimes_{i}^{N}
|\Psi_{\mathrm{block}}\rangle_{i,i-1}
\nonumber\\
&\propto
\bigotimes_{i}^{N}
\left[
\sum_{\bm{s}}
\hat U^{s,a}
\left(
|\bm{s}\rangle_{R_i}
|\bm a\rangle_{A_i}
\right)
\otimes
|\bm{s}\rangle_{L_{i-1}}
\right].
\label{eq:Phinew}
\end{align}
Eq.~\ref{eq:Phinew} retains the coupled-wire structure of the SPT, illustrated as Fig.~\ref{dualmain2}-c. It factorizes into the \textit{elementary blocks} of Eq.~\ref{eq:block}, each involving \(L_{i-1}\), \(R_i\), and the ancilla wire \(A_i\). Different blocks are decoupled, and stacking them along \(y\) produces the 2d wavefunction in Eq.~\ref{eq:Phinew}. Under the spacetime duality, \(L_{i-1}\), \(R_i\), and \(A_i\) are respectively the input, output, and ancilla of a single channel update. 

We now provide a diagrammatic tensor-network representation of the SPT wavefunction in Eq.~\ref{eq:Phinew}:
\begin{equation}
|\Phi\rangle =
\vcenter{\hbox{
\begin{tikzpicture}[x=1cm,y=1cm,baseline=(current bounding box.center)]
\node at (-2.35,0) {\(\cdots\)};
\node[coltensor] (A0) at (-1.35,0) {\(\mathcal A_{i-1}\)};
\node[coltensor] (A1) at (0,0) {\(\mathcal A_i\)};
\node[coltensor] (A2) at (1.35,0) {\(\mathcal A_{i+1}\)};
\node at (2.35,0) {\(\cdots\)};

\draw[tleg] (A0.east)--(A1.west);
\draw[tleg] (A1.east)--(A2.west);
\node[font=\scriptsize] at (.68,.22) {};

\draw[tleg] (-1.69,-.30)--(-1.69,-.78);
\draw[tleg] (-1.35,-.30)--(-1.35,-.78);
\draw[tleg] (-1.01,-.30)--(-1.01,-.78);

\draw[tleg] (-.34,-.30)--(-.34,-.78);
\draw[tleg] (0,-.30)--(0,-.78);
\draw[tleg] (.34,-.30)--(.34,-.78);
\node[font=\scriptsize] at (-.34,-.98) {\(L_i\)};
\node[font=\scriptsize] at (0,-.98) {\(R_i\)};
\node[font=\scriptsize] at (.34,-.98) {\(A_i\)};

\draw[tleg] (1.01,-.30)--(1.01,-.78);
\draw[tleg] (1.35,-.30)--(1.35,-.78);
\draw[tleg] (1.69,-.30)--(1.69,-.78);
\end{tikzpicture}
}}
\label{eq:pure_SPT_diagrammain}
\end{equation}
\begin{equation}
\vcenter{\hbox{
\begin{tikzpicture}[x=1cm,y=1cm,baseline=(current bounding box.center)]

\node[coltensor] (Ai) at (-2.2,-.15) {\(\mathcal A_i\)};

\draw[tleg] (Ai.west)--++(-.75,0);
\draw[tleg] (Ai.east)--++(.75,0);

\node[font=\scriptsize] at (-3.15,.08) {\(\nu_i\)};
\node[font=\scriptsize] at (-1.35,.08) {\(\mu_i\)};

\draw[tleg] (-2.50,.15)--(-2.50,.90);
\draw[tleg] (-2.20,.15)--(-2.20,.90);
\draw[tleg] (-1.90,.15)--(-1.90,.90);

\node[font=\scriptsize] at (-2.50,1.08) {\(L_i\)};
\node[font=\scriptsize] at (-2.20,1.08) {\(A_i\)};
\node[font=\scriptsize] at (-1.90,1.08) {\(R_i\)};

\node at (-.40,.20) {\(\displaystyle =\)};


\draw[tleg]
(1.20,.90)--(1.20,-.12)
.. controls (1.20,-.50) and (.82,-.50) ..
(.62,-.15)--(.25,-.15);

\draw[tleg] (1.75,.90)--(1.75,-.54);
\fill (1.75,-.54) circle (.075);

\draw[tleg]
(2.30,.90)--(2.30,-.12)
.. controls (2.30,-.50) and (2.68,-.50) ..
(2.88,-.15)--(3.25,-.15);

\node[font=\scriptsize] at (.43,.08) {\(\nu_i\)};
\node[font=\scriptsize] at (3.07,.08) {\(\mu_i\)};

\node[
    draw=blue,
    fill=cyan!65,
    text=black,
    minimum width=1.05cm,
    minimum height=.55cm,
    font=\large
] at (2.025,.40) {\(\hat U^{s,a}\)};

\node[font=\scriptsize] at (1.20,1.08) {\(L_i\)};
\node[font=\scriptsize] at (1.75,1.08) {\(A_i\)};
\node[font=\scriptsize] at (2.30,1.08) {\(R_i\)};

\end{tikzpicture}
}}
\label{eq:A_tensor_decomposition}
\end{equation}
In Eq.~\ref{eq:pure_SPT_diagrammain}, each MPS tensor $\mathcal{A}_i$ represents a physical column at $y=i$, whose physical Hilbert space consists of the $L_{i-1}$ and $R_i$ chains together with the ancilla chain $A_i$, represented by the corresponding physical legs. The virtual wires connecting neighboring Matrix Product State (MPS) tensors encode the coupled-wire structure. The detailed construction of the MPS tensor $\mathcal{A}_i$ and its explicit relation to the quantum channel are illustrated in Eq.~\ref{eq:A_tensor_decomposition}; further details are provided in Appendix~\ref{app:tensor_network} with field theory argument elucidated in Appendix~\ref{sec:wzw}.

We now give three equivalent interpretations of Eq.~\ref{eq:Phinew}.

\textbf{Interpretation I: pure-state SPT.}

One may start from the fixed-point SPT wavefunction \(|\Phi_0\rangle\) in Eq.~\ref{eq:Phi} consists of coupled EPR pair, adjoin a trivial ``ancilla layer'', and apply the symmetry-preserving local unitary \(\hat U^{s,a}\) that entangle them. This changes the microscopic wavefunction but not its SPT nature. Consequently, the pure state \(|\Phi\rangle\) (in Eq.~\ref{eq:Phinew}) defined in the enlarged Hilbert space remains a 2d SPT wavefunction protected by \(\mathrm{SO}(3)\times T_x\).

\textbf{Interpretation II: mixed-state SPT.}

Tracing out the ancilla layer after applying the unitary \(\hat U^{s,a}\) is equivalent to applying a local quantum channel that decoheres the fixed-point SPT state. This channel preserves the strong \(\mathrm{SO}(3)\) symmetry, while translation symmetry is preserved only weakly after the ancilla is traced out. The 2d reduced density matrix obtained after tracing out the ancilla from \(|\Phi\rangle\) (in Eq.~\ref{eq:Phinew}) is therefore
\begin{align}
\rho^{\mathrm{mSPT}}
=
\Tr_A
\left[
|\Phi\rangle\langle\Phi|
\right],
\label{eq:mSPT}
\end{align}
which realizes a 2d mSPT protected by strong \(\mathrm{SO}(3)\) symmetry and weak translation symmetry \(T_x\). 

\begin{equation}
\rho^{\mathrm{mSPT}}=
\vcenter{\hbox{
\begin{tikzpicture}[x=1cm,y=1cm,baseline=(current bounding box.center)]

\node[coltensor] (Km) at (-1.6,-.60) {\(\mathcal A_{i-1}\)};
\node[coltensor] (K)  at (0,-.60)    {\(\mathcal A_i\)};
\node[coltensor] (Kp) at (1.6,-.60)  {\(\mathcal A_{i+1}\)};

\node[coltensor] (Bm) at (-1.6,.60) {\(\mathcal A_{i-1}^{\dagger}\)};
\node[coltensor] (B)  at (0,.60)    {\(\mathcal A_i^{\dagger}\)};
\node[coltensor] (Bp) at (1.6,.60)  {\(\mathcal A_{i+1}^{\dagger}\)};

\draw[tleg] (Km.east)--(K.west);
\draw[tleg] (K.east)--(Kp.west);
\draw[tleg] (Bm.east)--(B.west);
\draw[tleg] (B.east)--(Bp.west);

\draw[tleg] (Km.west)--++(-.38,0);
\draw[tleg] (Bm.west)--++(-.38,0);
\draw[tleg] (Kp.east)--++(.38,0);
\draw[tleg] (Bp.east)--++(.38,0);

\draw[tleg] (Km.north east)--(Bm.south east);
\draw[tleg] (K.north east)--(B.south east);
\draw[tleg] (Kp.north east)--(Bp.south east);
\node[font=\scriptsize,right] at (.58,0) {\(\Tr_{A_i}\)};

\draw[tleg] (Km.235)--++(0,-.34);
\draw[tleg] (Km.265)--++(0,-.34);
\draw[tleg] (Bm.125)--++(0,.34);
\draw[tleg] (Bm.95)--++(0,.34);

\draw[tleg] (K.235)--++(0,-.34);
\draw[tleg] (K.265)--++(0,-.34);
\draw[tleg] (B.125)--++(0,.34);
\draw[tleg] (B.95)--++(0,.34);

\draw[tleg] (Kp.235)--++(0,-.34);
\draw[tleg] (Kp.265)--++(0,-.34);
\draw[tleg] (Bp.125)--++(0,.34);
\draw[tleg] (Bp.95)--++(0,.34);

\node[font=\scriptsize] at (-.14,-1.50) {\(L_i,\ R_i\)};

\end{tikzpicture}
}}
\label{eq:mSPT_diagrammain}
\end{equation}
Eq.~\ref{eq:mSPT_diagrammain} gives the diagrammatic representation of the two-dimensional mSPT density matrix in terms of its Matrix Product Operator (MPO) tensor representation. The top and bottom layers correspond to the ket and bra spaces of the density matrix, respectively, while the ancilla legs $A_i$ are contracted between the two ket/bra layers.

\textbf{Interpretation III: channel blocks.}

Under the spacetime rotation introduced in Fig.~\ref{dualmain2}-c, every entangled building block \(\ket{\Psi_{\mathrm{block}}}_{i,i-1}\) in Eq.~\ref{eq:Phinew} is the purification of one channel update: \(L_{i-1}\) is the input qubits, \(R_i\) is the output qubits, and \(A_i\) is the output ancilla. Thus, Eq.~\ref{eq:Phinew} represents a stack of disconnected channel along the \(y\) direction. 

As illustrated in Fig.~\ref{dualmain2}-d, projecting \(L_i\) and \(R_i\) in each bulk column onto onsite EPR pairs stitches the blocks together and connects the two spatial boundaries. Under spacetime rotation, each projected column corresponds connecting two adjacent channel steps. Thus, imposing these onsite-EPR projections across \(N\) bulk columns is equivalent to evolving the channel for \(N\) time steps. The initial and final state of the channel then become the spatial open boundaries of the 2d mSPT after bulk projection: the boundary at \(y=0\) carries the input state in its Choi-vectorized form \(|\rho_0\rangle\rangle\), while the boundary at \(y=N\) carries the output state \(|\mathcal{E}^N[\rho_0]\rangle\rangle\). We establish this correspondence explicitly in the next section.

\subsubsection{Repeating the quantum channel}\label{sec:renyi}

Building on the coupled-wire construction above, we now relate the repeated 1d quantum channel \(Q\), defined in Eq.~\ref{eq:qtransfer}, to the 2d SPT wavefunction \(|\Phi\rangle\) in Eq.~\ref{eq:Phinew} that includes both system and ancilla bits. We first represent the pure-state density matrix \(|\Phi\rangle\langle\Phi|\) in its Choi-double form,
\[
|\Phi_1\rangle\otimes|\Phi_2\rangle,
\]
where the two layers \(\alpha=1,2\) correspond to the ket and bra copies, respectively. As we had not tracing over the ancilla degrees of freedom, the state $|\Phi_1\rangle\otimes|\Phi_2\rangle$ is a pure bilayer SPT state. This should be distinguished from the mixed state \(|\rho^{\mathrm{mSPT}} \rangle\rangle\) in its Choi-double form, which arises only after the ancilla degrees of freedom are traced out.
Such a bilayer SPT state is protected by
\[
\mathrm{SO}(3)_a\times\mathrm{SO}(3)_b\times T_x.
\]
Here \(\mathrm{SO}(3)_a\) acts on the system spins in the first, ket layer, while \(\mathrm{SO}(3)_b\) acts on the system spins in the second, Hermitian-conjugate bra layer. The two \(\mathrm{SO}(3)\) actions are independent in the Choi-double Hilbert space as a manifestation of strong symmetry. The translation symmetry \(T_x\) acts simultaneously on the system and ancilla chains in both layers.

To begin with, we take the \(|\Phi_1\rangle\otimes|\Phi_2\rangle\) state and contract it with a `trivial reference state'. This gives
\begin{align}
\langle \phi^{\mathrm{tri}}|
\left(
|\Phi_1\rangle\otimes|\Phi_2\rangle
\right)
=
\mathrm{Tr}\!\left[Q^N\right].
\label{eq:relation}
\end{align}
The reference state is
\begin{align}
|\phi^{\mathrm{tri}}\rangle
\propto
\bigotimes_i
\left[
\bigotimes_{\alpha=1}^{2}
\left(
\sum_{\bm{s}}
|\bm{s}\rangle_{L_i}^{\alpha}
|\bm{s}\rangle_{R_i}^{\alpha}
\right)
\right]
\otimes
\left(
\sum_{\boldsymbol{\alpha}}
|\boldsymbol{\alpha}\rangle_{A_i}^{1}
|\boldsymbol{\alpha}\rangle_{A_i}^{2}
\right),
\end{align}
where \(\bm{s}\) and \(\boldsymbol{\alpha}\) label complete onsite-basis configurations of the system and ancilla chains, respectively.

Eq.~\ref{eq:relation} expresses the trace of the repeated channel as an overlap between a doubled 2d SPT wavefunction and a trivial reference state under periodic boundary conditions. The reference state contains two types of EPR pair. First, the interlayer EPR pairs
\(\sum_{\boldsymbol{\alpha}}
|\boldsymbol{\alpha}\rangle_{A_i}^{1}
|\boldsymbol{\alpha}\rangle_{A_i}^{2}\)
contract the ancilla degrees of freedom between the two layers. Second, within each layer, the onsite EPR pairs
\(\sum_{\bm{s}}
|\bm{s}\rangle_{L_i}^{\alpha}
|\bm{s}\rangle_{R_i}^{\alpha}\)
connect the \(L_i\) and \(R_i\) chains in the same column.
This projection has a direct channel interpretation. Within each layer, projecting \(L_i\) and \(R_i\) in the same column onto an EPR pair concatenates adjacent elementary blocks \(\ket{\Psi_{\mathrm{block}}}_{i,i-1}\). Repeating this contraction along the \(y\) direction generates the \(N\)-step channel \(Q^N\). The EPR projection over the interlayer ancilla degrees of freedom implements the trace over the ancilla qubits. Together, these contractions yield \(\mathrm{Tr}[Q^N]\).

Given that tracing out the ancilla degrees of freedom in \(|\Phi\rangle\) gives the 2d mixed state \(\rho^{\mathrm{mSPT}}\) defined in Eq.~\ref{eq:mSPT}. Eq.~\ref{eq:relation} can then be written as
\begin{align}
\langle \phi^{\mathrm{tri}}|
\left(
|\Phi_1\rangle\otimes|\Phi_2\rangle
\right)
=
\mathrm{Tr}[Q^N]
=
\mathrm{Tr}\!\left[
\rho^{\mathrm{mSPT}}\rho^{\mathrm{tri}}
\right].
\label{eq:mixed_overlap}
\end{align}
Here \(\rho^{\mathrm{tri}}\) is the density matrix of the trivial pure state formed by onsite EPR pairs between \(L_i\) and \(R_i\). Thus, the time transfer matrix of the repeated channel \(\mathrm{Tr}[Q^N]\) is mapped to the density matrix distance between the 2d mSPT and a trivial state.
One can also illustrate a diagrammatic representation of Eq.~\ref{eq:mixed_overlap}:
\begin{equation}
\mathrm{Tr}[Q^N]=
\vcenter{\hbox{
\begin{tikzpicture}[x=1cm,y=1cm,baseline=(current bounding box.center)]
\node[coltensor] (Km) at (-1.55,-.60) {\(\mathcal A_{i-1}\)};
\node[coltensor] (K)  at (0,-.60)     {\(\mathcal A_i\)};
\node[coltensor] (Kp) at (1.55,-.60)  {\(\mathcal A_{i+1}\)};

\node[coltensor] (Bm) at (-1.55,.60) {\(\mathcal A_{i-1}^\dagger\)};
\node[coltensor] (B)  at (0,.60)     {\(\mathcal A_i^\dagger\)};
\node[coltensor] (Bp) at (1.55,.60)  {\(\mathcal A_{i+1}^\dagger\)};

\node at (-2.45,-.60) {\(\cdots\)};
\node at (2.45,-.60)  {\(\cdots\)};
\node at (-2.45,.60)  {\(\cdots\)};
\node at (2.45,.60)   {\(\cdots\)};

\draw[tleg] (Km.west) -- ++(-.35,0);
\draw[tleg] (Km.east) -- (K.west);
\draw[tleg] (K.east)  -- (Kp.west);
\draw[tleg] (Kp.east) -- ++(.35,0);

\draw[tleg] (Bm.west) -- ++(-.35,0);
\draw[tleg] (Bm.east) -- (B.west);
\draw[tleg] (B.east)  -- (Bp.west);
\draw[tleg] (Bp.east) -- ++(.35,0);

\draw[tleg] (Km.55) -- (Bm.305);
\draw[tleg] (K.55)  -- (B.305);
\draw[tleg] (Kp.55) -- (Bp.305);
\path (K.55) -- (B.305) coordinate[pos=.5] (Atr);
\node[font=\scriptsize,right] at (Atr) {\(\Tr_A\)};

\path (Km.235) ++(0,-.28) coordinate (KmL);
\path (Km.265) ++(0,-.28) coordinate (KmR);
\draw[tleg] (Km.235) -- (KmL)
    .. controls +(0,-.18) and +(0,-.18) .. (KmR)
    -- (Km.265);

\path (K.235) ++(0,-.28) coordinate (KL);
\path (K.265) ++(0,-.28) coordinate (KR);
\draw[tleg] (K.235) -- (KL)
    .. controls +(0,-.18) and +(0,-.18) .. (KR)
    -- (K.265);

\path (Kp.235) ++(0,-.28) coordinate (KpL);
\path (Kp.265) ++(0,-.28) coordinate (KpR);
\draw[tleg] (Kp.235) -- (KpL)
    .. controls +(0,-.18) and +(0,-.18) .. (KpR)
    -- (Kp.265);

\path (Bm.125) ++(0,.28) coordinate (BmL);
\path (Bm.95)  ++(0,.28) coordinate (BmR);
\draw[tleg] (Bm.125) -- (BmL)
    .. controls +(0,.18) and +(0,.18) .. (BmR)
    -- (Bm.95);

\path (B.125) ++(0,.28) coordinate (BL);
\path (B.95)  ++(0,.28) coordinate (BR);
\draw[tleg] (B.125) -- (BL)
    .. controls +(0,.18) and +(0,.18) .. (BR)
    -- (B.95);

\path (Bp.125) ++(0,.28) coordinate (BpL);
\path (Bp.95)  ++(0,.28) coordinate (BpR);
\draw[tleg] (Bp.125) -- (BpL)
    .. controls +(0,.18) and +(0,.18) .. (BpR)
    -- (Bp.95);
\end{tikzpicture}
}}
\label{eq:standard_channel_diagrammain}
\end{equation}
 Since the mSPT density matrix admits an MPO representation, the onsite EPR projection between $L_i$ and $R_i$ component is implemented by contracting the two physical tensor legs $L_i$ and $R_i$ within the same MPO tensor as Eq.~\ref{eq:standard_channel_diagrammain}. Under spacetime duality, this contraction corresponds to connecting successive quantum channels.

We have thus shown that the overlap between the bilayer SPT state and a trivial reference state is directly related to the trace of the repeated channel \(\mathrm{Tr}[Q^N]\). Importantly, the reference state \(|\phi^{\mathrm{tri}}\rangle\) preserves the same
\(\mathrm{SO}(3)_a\times\mathrm{SO}(3)_b\times T_x\) symmetry as the Choi-double state. This overlap underlies the strange-correlator construction for the SPT state, which we discuss in the next section.

\color{black}

\subsubsection{ Twisted Rényi-$N$ correlator}\label{sec:prooftrnc}

\textbullet{~\textbf{Twisted Rényi-\(N\) correlator for strong symmetry:}}

\begin{figure}[h]
    \centering
\includegraphics[width=0.6\linewidth] {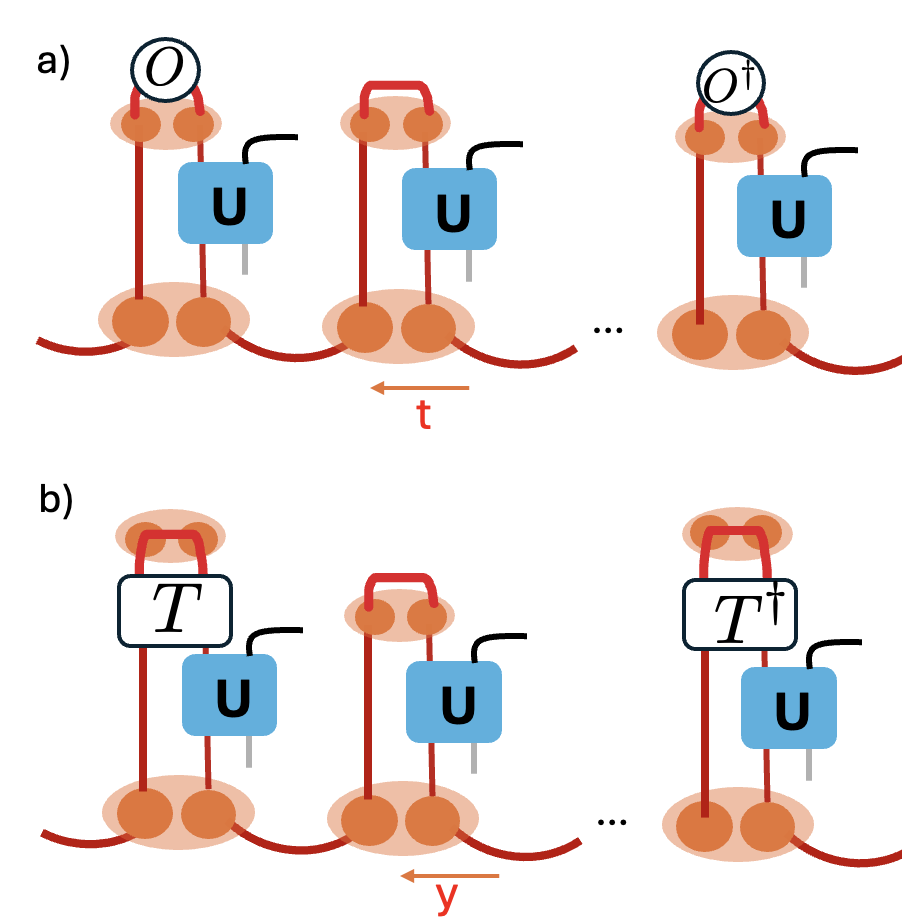}
    \caption{a) The TRNC of  Q probes temporal correlations generated by the repeated quantum channel.
b) After implementing spacetime duality, the TRNC is mapped to the strange correlator of the corresponding 2d SPT state.}
    \label{scfig}
\end{figure}

\begin{figure}[h]
    \centering
    \includegraphics[width=0.3\textwidth]{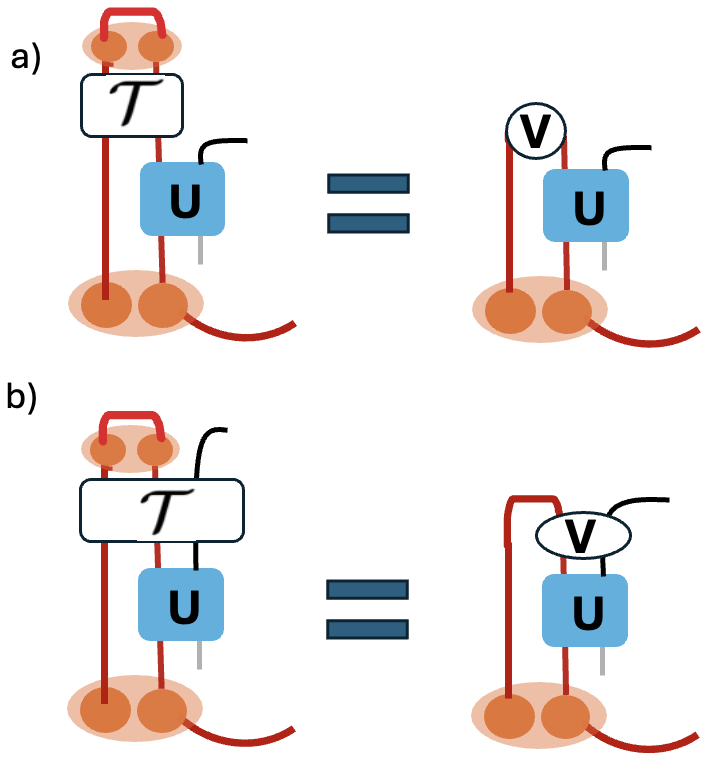}
    \caption{ a) Operator mapping for the TRNC in Eq.~\ref{eq:twsited1}, which detects temporal correlations of the strong-symmetry charge. The left panel shows the corresponding operator insertion in the 2d SPT, while the right panel shows how the operators are inserted in the 1d quantum channel.
b) Operator mapping for the TRNC in Eq.~\ref{eq:twsitedw}, which detects temporal correlations of the weak-symmetry charge. The left panel shows the corresponding operator insertion in the 2d SPT, while the right panel shows how the operators are inserted in the quantum channel. }
    \label{fig:strongweakope}
\end{figure}

The preceding construction maps the repeated channel \(Q\) to the bilayer
SPT state \(|\Phi_1\rangle\otimes|\Phi_2\rangle\). Since this state is a
2d SPT, its strange correlators~\cite{zhang2022strange,lepori2023strange}
are expected to exhibit long-range or quasi-long-range order along the \(y\) direction.
We first consider a strange correlator charged under
\(\mathrm{SO}(3)_a\), or equivalently under \(\mathrm{SO}(3)_b\).
As illustrated in Fig.~\ref{scfig}, spacetime duality maps this correlator
to the twisted Rényi-\(N\) correlator (TRNC) of the 1d quantum channel \(Q\)
in Eq.~\ref{eq:autoco}:
\begin{align}
&C^{s}(N)
=
\frac{
\langle\phi^{\mathrm{tri}}|
T^{\dagger}(x,0)T(x,N)
|\Phi_1\otimes\Phi_2\rangle
}{
\langle\phi^{\mathrm{tri}}|
\Phi_1\otimes\Phi_2\rangle
}
\nonumber\\
&=
\frac{
\mathrm{Tr}\!\left[
O^{\dagger}(x)Q^{N}O(x)Q^{N}
\right]
}{
\mathrm{Tr}\!\left[Q^{2N}\right]
},
\label{eq:twsited1}\\
T&=\mathcal{T}_1\otimes\mathcal{T}_2^{\dagger},
\qquad
O=V_1\otimes V_2^{\dagger},
\nonumber\\
&\langle c_L,d_R|\mathcal{T}|a_L,b_R\rangle\,
\delta_{cd}
=
\langle a|V|b\rangle .
\nonumber
\end{align}
The first line is the \textit{strange correlator} of the 2d SPT state, in which \(\mathrm{SO}(3)_a\)-charged operators are inserted at two positions along the \(y\) direction and their spatial correlation is evaluated. Under the spacetime-duality mapping, the second line becomes the \textit{Twisted Renyi-N Correlator (TRNC)} of the repeated quantum channel: the strong  \(\mathrm{SO}(3)_a\)-charged operators are inserted between different time intervals, and their temporal correlator is measured (See Appendix \ref{sec:phys} for more discussions).
Here, \(\mathcal{T}_{i}\) are local spin operators acting in the $i=1,2$
layer. The bilayer operator
\(T=\mathcal{T}_1\otimes\mathcal{T}_2^\dagger\) is charged under both
\(\mathrm{SO}(3)_a\) and \(\mathrm{SO}(3)_b\), while it is neutral under
their diagonal action. In the channel representation,
\(O=V_1\otimes V_2^\dagger\) acts on the ket and bra spaces, with \(V\)
carrying nontrivial \(\mathrm{SO}(3)\) charge.
The operator \(\mathcal{T}\) acts on both the \(L\) and \(R\) chains within
one column of the 2d wavefunction, whereas \(V\) acts on the single physical
chain of the 1d channel. Their tensor-element correspondence is given in
Eq.~\ref{eq:twsited1} and illustrated in
Fig.~\ref{fig:strongweakope}-a. The second line of
Eq.~\ref{eq:twsited1} is therefore the TRNC
of the transfer matrix \(Q\) along the time direction
\cite{sun2024holographic,sala2025entanglement}.

Eq.~\ref{eq:twsited1} can equivalently be expressed as the type-II
strange correlator of the 2d mSPT in Eq.~\ref{eq:mSPT}, introduced in
Refs.~\cite{zhang2022strange,lee2025symmetry} as a mixed-state SPT
diagnostic:
\begin{align}
&
\frac{
\langle\phi^{\mathrm{tri}}|
T^{\dagger}(x,0)T(x,N)
|\Phi_1\otimes\Phi_2\rangle
}{
\langle\phi^{\mathrm{tri}}|
\Phi_1\otimes\Phi_2\rangle
}
\nonumber\\
&\quad =
\frac{
\mathrm{Tr}\!\big[
\mathcal{T}(x,0)\mathcal{T}^{\dagger}(x,N)\,
\rho^{\mathrm{mSPT}}\,
\mathcal{T}^{\dagger}(x,0)\mathcal{T}(x,N)\,
\rho^{\mathrm{tri}}
\big]
}{
\mathrm{Tr}\!\big[
\rho^{\mathrm{mSPT}}\rho^{\mathrm{tri}}
\big]
}.
\end{align}

\textbullet{{~\textbf{ Twisted Rényi-\(N\) correlator for weak symmetry:}}

We next consider the strange correlator charged under the weak translation
symmetry \(T_x\). In contrast to the strong-symmetry case, translation acts
on both the system and ancilla degrees of freedom in each layer. The charged
operator can therefore act jointly on these two degrees of freedom:
\begin{align}
C^{w}(N)
&=
\frac{
\langle\phi^{\mathrm{tri}}|
T^{\dagger}(x,0)T(x,N)
|\Phi_1\otimes\Phi_2\rangle
}{
\langle\phi^{\mathrm{tri}}|
\Phi_1\otimes\Phi_2\rangle
}
\nonumber\\
&=
\frac{
\mathrm{Tr}\!\left[
\tilde{Q}^{\dagger}(x)
Q^{N-1}
\tilde{Q}(x)
Q^{N-1}
\right]
}{
\mathrm{Tr}\!\left[Q^{2N}\right]
}.
\label{eq:twsitedw}
\end{align}
The operator mapping, illustrated in Fig.~\ref{fig:strongweakope}-b, is
\begin{align}
T&=\mathcal{T}_1\otimes I_2,
\qquad
\langle j,c_L,d_R|\mathcal{T}|i,a_L,b_R\rangle\,
\delta_{cd}
=
\langle j,a|V|i,b\rangle,
\nonumber\\
&\sum_{c,i}
\langle j,a|V|i,c\rangle
\langle c|K_i|b\rangle
=
\langle a|\tilde{K}_j|b\rangle,
\nonumber\\
&\sum_j K_j\otimes K_j^{*}
=Q,
\qquad
\sum_j\tilde{K}_j\otimes\tilde{K}_j^{*}
=\tilde{Q}.
\label{eq:explain}
\end{align}
The operator \(T=\mathcal{T}_1\otimes I_2\) acts only in the first layer, as appropriate for a weak
symmetry insertion. Within that layer, however, \(\mathcal{T}_1\) acts on the
ancilla and on the \(L\) and \(R\) chains. Its tensor elements are therefore
labeled by \((i,a_L,b_R)\), where \(i\) labels the ancilla state and
\(a_L,b_R\) label the two system legs.

Under spacetime duality, \(C^{w}(N)\) becomes the TRNC probing temporal
correlations of the weak \(T_x\) charge. Because the weak-symmetry operator
acts jointly on the system and ancilla, its insertion modifies the Kraus
operators from \(K_j\) to \(\tilde{K}_j\). Accordingly,
\(\tilde{Q}\) is the channel transfer matrix with a momentum insertion, as
specified in Eq.~\ref{eq:explain}.

For SPT wavefunctions,
Refs.~\cite{you2014wave,zhang2022strange,lee2025symmetry} showed that the
corresponding strange correlators must exhibit either long-range or
quasi-long-range order. In particular, for the bilayer SPT state 
\(|\Phi_1\rangle\otimes|\Phi_2\rangle\) protected by
\(\mathrm{SO}(3)_a\times\mathrm{SO}(3)_b\times T_x\), at least one strange
correlator charged under an \(\mathrm{SO}(3)\) symmetry or under \(T_x\) must
display (quasi-) long-range order\footnote{In some cases, both correlators
exhibit quasi-long-range order.}. Since this strange correlator along the
\(y\) direction maps to a TRNC of the quantum channel, at least one of
\(C^{s}(N)\) in Eq.~\ref{eq:twsited1} and \(C^{w}(N)\) in
Eq.~\ref{eq:twsitedw} must exhibit (quasi-)long-range temporal behavior.

\emph{Final Remark}:
The long-range order of the TRNC immediately implies that the spectrum of $Q$ contains either gapless or degenerate modes. However, a gapless or degenerate Liouvillian spectrum is not, by itself, a distinctive signature of the LSM constraint, since such behavior already arises generically in strongly symmetric quantum channels. A prominent example is a channel with strong $U(1)$ symmetry, where diffusive modes guarantee gapless eigenvalues and consequently long mixing times~\cite{huang2024hydro,gu2024spontaneous,hauser2026strong}. Meanwhile, for an anomaly-free channel, a featureless steady state can coexist with a gapless Liouvillian. Specifically, even when the Liouvillian spectrum is gapless, the dynamics may relax to a steady state with finite correlation length and finite Markov length. Thus, the featurelessness of the steady state, as diagnosed by ordinary correlation functions and fidelity correlators, is not in general related to the Liouvillian gap. This is sharply different from the equilibrium LSM setting, where a gapless Hamiltonian is typically accompanied by long-range or algebraic correlations in the ground state. We discuss this distinction in detail in Appendix~\ref{app:featureless}.

Motivated by the above observation, in the next section, we introduce a different diagnostic: \textit{The Liouvillian singular spectrum}, which is defined as \textit{the spectrum of $\Omega = Q^\dagger Q$}. We will show that the LSM constraint enforces $\Omega$ to be gapless or degenerate. In addition, we show that the TRNC associated with $\Omega$ also exhibits long-range order. Together, these results suggest that $\Omega$ provides a more faithful probe of the LSM constraint and its associated anomaly.

\section{Singular Spectrum of $Q$}\label{sec:Sing}

Beyond the spectral properties of the quantum channel’s transfer matrix, it is worthwhile to elaborate on several additional salient features of $Q$. 
When analyzing a non-Hermitian operator such as our transfer matrix $Q$, it is instructive to consider not only its right eigenvectors and eigenvalues, but also its singular value decomposition (SVD) and associated singular value spectrum. In particular, one can define a Hermitian operator $\Omega =  Q^{\dagger} Q$, whose eigenvalues correspond to the squared singular values of $Q$.

\begin{figure}[h]
    \centering
\includegraphics[width=0.4\textwidth]{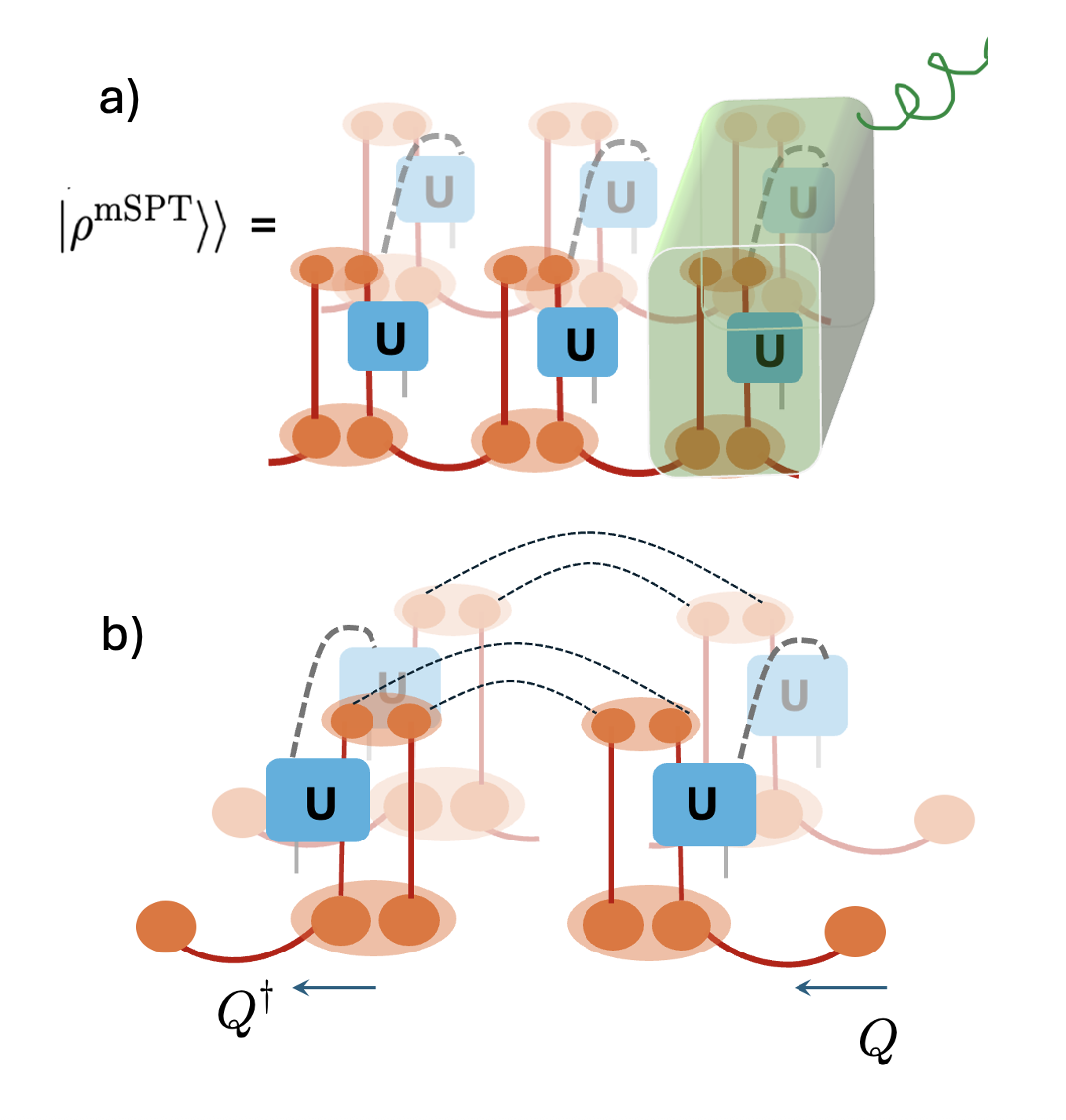}
    \caption{ a) The 2d mSPT  defined in Eq.~\ref{eq:mSPT}, shown in its Choi-doubled representation. The front and back layers correspond to the ket and bra spaces of $\rho^{\mathrm{mSPT}}$, respectively.
 b) A spatial bipartition of $|\rho^{\mathrm{mSPT}}\rangle\rangle$ yields a reduced density matrix (RDM). Owing to the coupled-wire structure, it is sufficient to analyze a single building block of the wavefunction, whose RDM is related to $\Omega=Q^{\dagger}Q$. }
    \label{fig:wavefun}
\end{figure}

Starting from the 2d mSPT density matrix in Eq.~\ref{eq:mSPT}, we represent it in the Choi-double form shown in Fig.~\ref{fig:wavefun}-a. It then follows that the reduced density matrix of $|\rho^{\mathrm{mSPT}}\rangle\rangle$ across a spatial cut is
\[
\rho^{\mathrm{RDM}}=\frac{\Omega}{\Tr[\Omega]},
\]
as illustrated in Fig.~\ref{fig:wavefun}-b.

Hence, the entanglement spectrum of $|\rho^{\mathrm{mSPT}}\rangle\rangle$ directly coincides with the eigenvalue spectrum of $\Omega$ (up to normalization). In this way, the Liouvillian singular spectrum is mapped to the \textit{operator entanglement} structure of the corresponding mSPT density matrix.

\begin{shaded}
\textbullet~ Lemma I: Let $Q$ denote the transfer matrix associated with a 1d quantum channel in superoperator space with weak symmetry $G$ and a strong symmetry $S$ that exhibit quantum anomaly. Then $\Omega \equiv  Q^{\dagger} Q$ is equivalent to the reduced density matrix of a 2d SPT defined in the Choi-double space, protected by $S_a \otimes S_b \otimes G$, with $S_a$ and $S_b$ acting independently on the ket and bra layers. The twisted Rényi-$N$ correlator (TRNC) with respect to $\Omega = Q^{\dagger} Q$ defined in the following exhibit quasi-long-range or long-range order along the replica direction:
\begin{align}\label{eq:twsited2}
    C_{\Omega}(N) = \frac{\Tr[ O^{\dagger}(x) \Omega^{N} O(x)\Omega^{N}]}{\Tr[\Omega^{2N}]}.
\end{align} 
\end{shaded}

Lemma I follows from the fact that $\Omega$ can be interpreted as the \textit{spatial reduced density matrix} of the 2d SPT wavefunction $|\rho^{\mathrm{mSPT}}\rangle\rangle$, defined in the doubled Hilbert space. As shown in Ref.~\cite{sala2025entanglement}, for the reduced density matrix of any SPT state, its TRNC defined as Eq.~\ref{eq:twsited2} necessarily exhibits a (quasi) long-range order. The proof, detailed in Ref.~\cite{sala2025entanglement}, proceeds by interpreting replicas of the reduced density matrix as a coupled-wire construction of the SPT fixed-point wavefunction and demonstrating that the TRNC defined for the reduced density matrix is equivalent to a strange correlator of the fixed-point SPT state.

As Eq.~\ref{eq:twsited2} exhibits long-range order along the replica direction, $\Omega$ cannot have a unique dominant eigenvector separated by a finite gap. The reasoning is as follows: suppose $\Omega$ has a unique gapped eigenvector $|\chi \rangle$ such that $\Omega^N \sim |\chi \rangle \langle \chi|$ becomes a projection operator. Since $|\chi \rangle$ is the unique dominant eigenvector, it must also be an eigenstate of the strong $S$ symmetry and the weak $G$ symmetry. If $O$ is a charged operator, then $O|\chi \rangle$ carries a different quantum number than $|\chi \rangle$. Consequently, $O|\chi \rangle$ is orthogonal to $|\chi \rangle$, and hence $C_{\Omega}(N)=0$.
This, in turn, implies that the spectrum of $\Omega$ must be gapless or degenerate. Thus, we ran into the following Proposition:

\begin{shaded}
\textbullet~ Proposition II: Let $Q$ be the transfer matrix of a quantum channel subject to the LSM constraint. Then the Liouvillian singular spectrum, or equivalently the eigenvalue spectrum of $\Omega \equiv Q^\dagger Q$, must be gapless or degenerate.
\end{shaded}
This proposition states that the LSM constraint forces the Liouvillian singular spectrum to be either gapless or degenerate. This result  follows naturally from the fact that the Liouvillian singular spectrum is analogous to the entanglement spectrum of a higher-dimensional SPT state, which cannot possess a unique, gapped dominant eigenvector.

Here and throughout, we distinguish between the \emph{Liouvillian spectrum}, defined as the eigenvalue spectrum of the transfer matrix \(Q\), and the \emph{Liouvillian singular spectrum}, defined as the spectrum of
\(\Omega=Q^{\dagger}Q\). Notably, the anomaly constrains the Liouvillian singular spectrum rather than the Liouvillian spectrum itself. The anomaly-enforced degeneracy or gaplessness of \(\Omega\) should therefore be distinguished from any degeneracy or gaplessness in the eigenvalue spectrum of \(Q\). Indeed, the latter can arise for reasons entirely unrelated to an anomaly. For a CPTP channel with a strong symmetry, \(Q\) decomposes into distinct strong-symmetry sectors, each supporting a steady state, and hence generally exhibits degeneracy. Likewise, in the presence of a strong \(U(1)\) symmetry, the Liouvillian spectrum of \(Q\) is generically gapless because of hydrodynamic modes, even in an anomaly-free system. Consequently, neither degeneracy nor gaplessness in the Liouvillian spectrum of \(Q\) is, by itself, an anomaly diagnostic. The relevant diagnostic is instead the anomaly-enforced structure of the Liouvillian singular spectrum encoded in \(\Omega\).

The spectrum of \(Q\) and its singular spectrum, can exhibit qualitatively different behavior. For example, an anomaly-free quantum channel with strong \(U(1)\) symmetry may have a gapless \(Q\) spectrum due to hydrodynamic modes, while the \(\Omega\) spectrum remains nondegenerate and gapped. In Appendix~\ref{sec:U1gapped}, we illustrate this distinction by constructing a translation-invariant integer-spin chain with strong \(U(1)\) symmetry for which \(\Omega\) has a unique, gapped dominant eigenvalue. Similarly, for a quantum channel with strong \(Z_2\) symmetry but no LSM constraint, the spectrum of \(Q\) is necessarily degenerate because of the strong-symmetry sectors, whereas \(\Omega\) may still possess a unique dominant eigenvalue. An explicit example is provided in Appendix~\ref{sec:nondeg}.

\subsection{Numerical evidence}\label{sec:num}

As an explicit illustration of our theory, we study a brickwork quantum channel on a periodic spin-$\tfrac{1}{2}$ chain of even length $L$. The channel always preserves the strong on-site $U(1)$ symmetry, while the lattice translation $T_x$ and the $Z_2$ spin-flip symmetry may be either preserved as weak symmetries or explicitly broken. The relevant internal symmetry operations are
\begin{align}\label{eq:sym1}
U(1): e^{i\phi \sum_j \sigma^j_z}, \qquad
Z_2: \prod_j \sigma^j_x .
\end{align}
When both $T_x$ and the $Z_2$ spin-flip symmetry are present as weak symmetries, their interplay with the strong $U(1)$ symmetry exhibits a mixed-state 't Hooft anomaly.

In our setting, one timestep consists of nearest-neighbor channels applied successively to the even and odd bonds. Each two-site channel first dephases the computational basis and then implements a classical stochastic update. Writing $\nu$ for the input state (column) and $\mu$ for the output state (row), the transition matrix is
\begin{equation}\label{eq:wmatrix}
W = \begin{pNiceMatrix}[first-row,last-col]
 \ket{00}   & \ket{01} & \ket{10} & \ket{11} &        \\
1   & 0   & 0   & 0   & \ \ket{00} \\
0   & a   & c   & 0   & \ \ket{01} \\
0   & b   & d   & 0   & \ \ket{10} \\
0   & 0 & 0 & 1 & \ \ket{11}       \\
\end{pNiceMatrix},
\end{equation}
with $a,b,c,d\ge 0$ and $a+b=c+d=1$. The local two-site quantum channel in the Choi-doubled space is given by
\begin{align}
Q_{\rm loc}&=
\sum_{\mu,\nu} W_{\mu\nu}
|\mu\rangle|\mu^*\rangle\langle\nu|\langle \nu^*|.
\end{align}
The full transfer matrix $Q$ is obtained by composing these local channels in the same even-odd brickwork pattern.
Due to this alternating structure, the dynamics generally do not possess an exact weak translation symmetry. However, we argue below that weak translation symmetry can emerge asymptotically at long times.

For arbitrary parameters satisfying these constraints, the local channel possesses a \textit{strong $U(1)$ symmetry}. To make the resulting dynamics also compatible with the \textit{weak $Z_2$ symmetry}, we can choose
$a=b=c=d=\frac12$.
In this case, the local channel is weakly symmetric under global spin flip. While an individual two-site channel does not possess weak translation symmetry $T_x$, the full dynamics generated by alternating even- and odd-bond layers does so approximately at long times.
To see this, note that the resulting dynamics is equivalent to an interacting random-walk process~\cite{liu_iso:2024,boesl_iso:2025,boesl:2026}, whose steady states are uniform classical mixtures over all product states within each fixed $U(1)$-symmetry sector and are therefore weakly translation invariant. Denoting the two-layer channel by $\Lambda$, contractivity implies that, for sufficiently large $l$, $\Lambda^l$ approaches the projector onto this steady-state manifold. Thus, at long times, $\Lambda^l$ approximately projects onto a weakly translation-invariant subspace, and the dynamics effectively realizes weak $T_x$ symmetry.


To demonstrate the unique feature of the LSM constraint,
we evaluate the connected TRNC $
\frac{\Tr[O^\dagger\Omega^N O\Omega^N]}{\Tr[\Omega^{2N}]}
-\left(\frac{\Tr[O\Omega^N]}{\Tr[\Omega^N]}\right)^2$ of $\Omega$ using $O=S_z^i\otimes I$. Here, $O$ carries charge under the weak symmetry and thus has a vanishing expectation value, so the connected TRNC reduces to the TRNC defined in Eq.~\ref{eq:twsited2}. As shown on a log-log scale in Fig.~\ref{fig:trnc_plot}-a, it exhibits behavior consistent with power-law decay as the system size increases, while the spectrum of $\Omega$ becomes gapless, as shown in Fig.~\ref{fig:twsited}-a.
We next explicitly break the weak $T_x \times Z_2$ symmetry by choosing $a = 0.04$ and $c = 0.36$, rendering the channel dynamics anomaly-free. In this case, $O$ is no longer charged under any symmetry and can acquire a nonzero expectation value, which is subtracted in the connected TRNC.
We find that the spectrum of $\Omega$ becomes gapped (Fig.~\ref{fig:twsited}-b), and the connected TRNC for $\Omega$ decays exponentially along the replica direction $N$, as shown in the semi-log plot in Fig.~\ref{fig:trnc_plot}-b.


 \begin{figure}[t]
     \centering
     \includegraphics[width=0.8\linewidth]{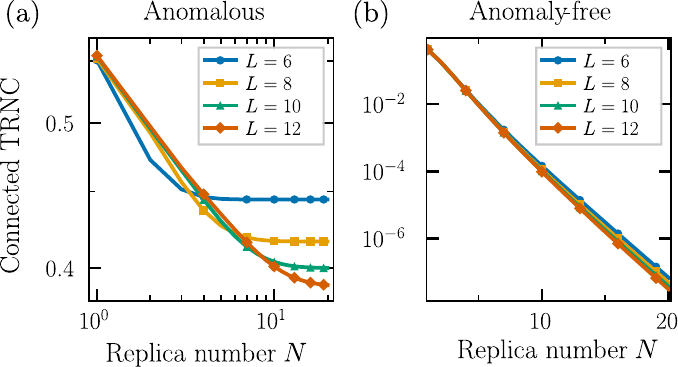}
     \caption{The connected TRNC constructed from $\Omega$ for quantum channels on rings of length $L=6,8,10,$ and $12$, plotted as a function of the replica number $N$. a) The connected TRNC of $\Omega$ for anomalous quantum-channel dynamics with $a=c=1/2$ is plotted in log-log scale. b) The connected TRNC for anomaly-free quantum-channel dynamics with $a=0.04$ and $c=0.36$, and with $y$-axis shown in log scale, demonstrating the exponential decay of the TRNC.} 
     \label{fig:trnc_plot}
 \end{figure}

As $\Omega$ is the RDM of a higher-dimensional SPT, any property of the RDM in the 2d SPT state transfers to $\Omega$. 
Here, we highlight a twisted observable, first proposed in Refs.~\cite{xu2025diagnosing, zang2024detecting}, which detects the underlying anomaly in $\Omega$ by probing its response to symmetry twists. Consider the transfer matrix $Q(\theta_1)$ that has a twisted boundary condition on the first layer (which is analogous to flux insertion with respect to the strong U(1) symmetry) and define:
\begin{align}
\Omega(\theta) = Q^{\dagger}(\theta)Q(\theta).
\end{align}
 The operator $\Omega(\theta)$ can be interpreted as an RDM of an SPT state with a twisted boundary condition parametrized by $\theta$. For a finite system of length $L$, one may adiabatically tune $\theta$ from $0$ to $2\pi$ and track the spectrum flow of $\Omega(\theta)$; a characteristic signature is a level crossing that occurs during this flux-threading cycle. Physically, varying $\theta$ implements flux insertion in the effective 1d theory, leading to a momentum shift in the spectrum. 

Here, we simulate the spectral flow of $\Omega(\theta)$ for the quantum channel defined in Eq.~\ref{eq:wmatrix}. The spectrum of $\Omega(\theta)$ is computed for the anomalous and anomaly-free parameter choices, $a=0.04$ and $c=0.36$, respectively. The results are shown in Fig.~\ref{fig:twsited}. In the anomalous case, shown in Fig.~\ref{fig:twsited}-a, there is a level crossing between the dominant and subdominant eigenvalues, indicating that the corresponding $\Omega$ spectrum is gapless. By contrast, in the anomaly-free case, shown in Fig.~\ref{fig:twsited}-b, the spectrum remains gapped, and the dominant eigenvector is unchanged under flux insertion.

 \begin{figure}[h]
     \centering
     \includegraphics[width=1\linewidth]{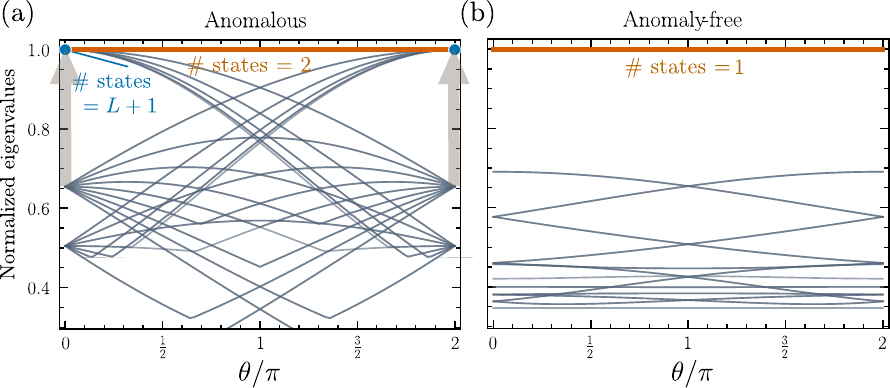}
     \caption{Spectrum of $\Omega(\theta)$, where $Q$ is defined with the same set of parameters in Fig.~\ref{fig:trnc_plot} in the anomalous and anomaly-free case. The largest 50 eigenvalues (normalized against the largest eigenvalues) for $L = 10$. (a) At $\theta = 0$ or $2\pi$ (blue dots), we recover the untwisted $\Omega$ with $(L+1)$-fold degenerate dominant eigenvalues and.gapless spectrum as $L\to\infty$ (grey arrows). The gaplessness is verified by varying the system size from $L=4$ to $L=12$ and observing a consistent closing of the gap to the dominant eigenvalues (data not shown). The two degenerate dominant eigenstates (orange line) are not responsive to the twist, they correspond to the fully occupied and the completely empty state. (b) The anomaly-free $U(1)$ symmetric example. The dominant eigenvalue remains nondegenerate. When varying from $L=4$ to $L=12$, the gap to the dominant eigenvalue of $\Omega$ appears to have little dependence on $L$ for all values of $\theta$.}
     \label{fig:twsited}
 \end{figure}

 \subsection{Steady state under LSM constraint}
 
We now discuss the implication of an LSM constraint for the steady state of a quantum channel. As initially explored in Refs.~\cite{lessa2025mixed, you2024intrinsic, xu2025average,lee2023quantum}, a quantum anomaly obstructs a featureless mixed state with both finite correlation length and finite Markov length. Here, we give a simple field-theoretic picture that makes this obstruction particularly transparent for the steady state generated by repeated application of a quantum channel.
As is detailed in Appendix~\ref{sec:wzw}, the repeated channel $Q^N$ can be represented as a Euclidean path integral in the Choi-doubled space, where the channel-evolution direction becomes Euclidean time $\tau\in[0,N]$. Thus $Q$ acts as a transfer matrix in the Choi-doubled superoperator space, with $N$ playing the role of an inverse temperature.
In the long-time limit $N\to\infty$, the channel projects onto its steady state. Written as a Choi-doubled vector $|\rho_{\mathrm{ss}}\rangle\rangle$, this steady state is analogous to the ground state of the corresponding Euclidean path integral.

The LSM constraint and its associated anomaly therefore obstruct a featureless steady state. Since the Euclidean theory in the Choi-doubled space is anomalous, its ground state ($|\rho_{\mathrm{ss}}\rangle\rangle$) cannot be both short-range entangled and symmetry preserving. Thus, within a fixed strong-symmetry sector, suppose the steady state is also invariant under weak symmetry, the steady-statete density matrix cannot be completely featureless. It must retain some form of long-range order in the Choi-doubled space: either SWSSB, diagnosed by nonvanishing CMI, or spontaneous breaking of the weak symmetry, diagnosed by nonvanishing MI~\cite{lessa2025mixed}.
An explicit example of the first scenario is the anomalous channel discussed in Sec.~\ref{sec:num} with $a=c=1/2$ in Eq.~\ref{eq:wmatrix}. The dynamics reduce to a one-dimensional random walk, so that the $S_z=0$ sector has a unique steady state given by the equal-weight mixture of all bitstrings in the sector. This state exhibits $U(1)$ SWSSB and has logarithmic operator entanglement scaling as $O(\log L)$. Notably, for generic quantum channels~\cite{lu2025holographic,ziereis2025strong}, SWSSB is expected to occur in the steady state. Nevertheless, fine-tuned quantum channel can produce steady states that break only the weak symmetry.
In Appendix~\ref{app:weakbreak}, we construct an absorbing channel on a spin-$\frac12$ chain that preserves both the strong on-site $U(1)$ symmetry and the weak $T_x\times Z_2$ symmetry defined in Sec.~\ref{sec:num}. In the $S_z=0$ sector, the steady state does not exhibit $U(1)$ SWSSB. Instead, it develops Néel order and spontaneously breaks the weak $T_x\times Z_2$ symmetry.

By contrast, an anomaly-free quantum channel can support a completely featureless steady state. One immediate example is obtained by setting $a=c=0$ in Eq.~\ref{eq:wmatrix}. The resulting channel retains only the strong $U(1)$ symmetry, while the weak $Z_2$ and translation symmetries are completely broken; consequently, the channel is anomaly-free. The dynamics are absorbing: the system remains unchanged unless an even bond is in the state $\ket{01}$, in which case it is deterministically mapped to $\ket{10}$. If the first circuit layer acts on the even bonds, the steady state in the $S_z=0$ sector is unique and is simply the trivial product state $\ket{10}^{\otimes L/2}$, which has no operator entanglement. See Appendix~\ref{app:featureless} for a translation-invariant example.

\subsection{Physical interpretation of $\Omega$} \label{sec:epr}

\begin{figure}[h]
    \centering
\includegraphics[width=0.25\textwidth]{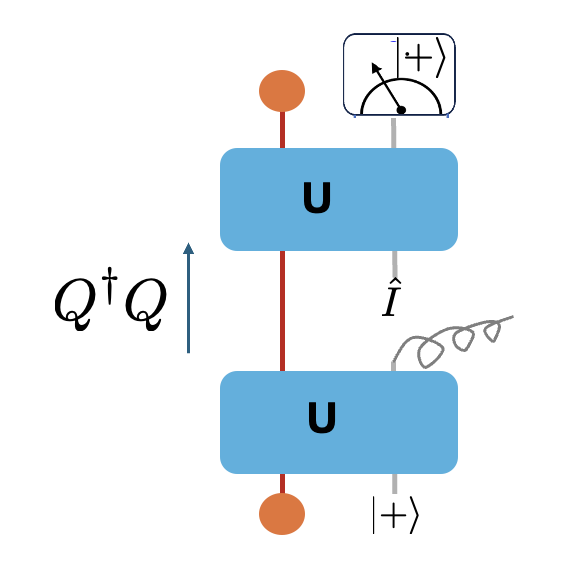}
    \caption{Schematic of $Q^\dagger Q$. $Q$ appends ancillas in tensor product state $\otimes|+\rangle$, entangles them with the system, and traces them out, while $Q^\dagger$ appends maximally mixed ancillas, applies the same unitary, and projects them onto $\otimes|+\rangle$.}
    \label{fig:omega}
\end{figure}

While we have shown that \(\Omega = Q^{\dagger}Q\) exhibits long-range order along the replica direction, its implication to physical dynamics is less direct. Recall that \(Q\) is the transfer matrix of the quantum channel in superoperator space. This channel admits a unitary purification: one introduces an ancilla layer initialized in a tensor product state \(\otimes \lvert + \rangle\) state that is fully polarized in some local basis, applies a unitary that entangles the system and ancilla, and subsequently traces out the ancilla to recover the original quantum channel.

Within the same construction, \(Q^{\dagger}\) represents the adjoint transfer matrix. Diagrammatically, it is obtained by appending an ancilla layer initialized in the maximally mixed state, represented by \(\hat I\), applying the corresponding unitary circuit that entangles the system and ancilla, and finally projecting the ancilla onto the onsite basis as \(\otimes \lvert + \rangle\). Owing to this final projection, the resulting map is completely positive but not trace-preserving.

As illustrated in Fig.~\ref{fig:omega}-a, one may regard \(\Omega\) as the transfer matrix of a repeated protocol that alternates the physical forward channel \(Q\) with the post-selected map \(Q^{\dagger}\). After \(N\) repetitions, the effective transfer matrix is
\begin{equation}\label{eq:omega}
\Omega^{N} = (Q^{\dagger}Q)^{N}.
\end{equation}
The twisted Rényi-\(N\) correlator for \(\Omega\), defined in Eq.~\ref{eq:twsited2}, therefore probes temporal correlations of this interleaved evolution, which combines ordinary open-system dynamics under \(Q\) with syndrome measurement and post-selection encoded by \(Q^{\dagger}\).

\begin{figure}[h]
    \centering
\includegraphics[width=0.5\textwidth]{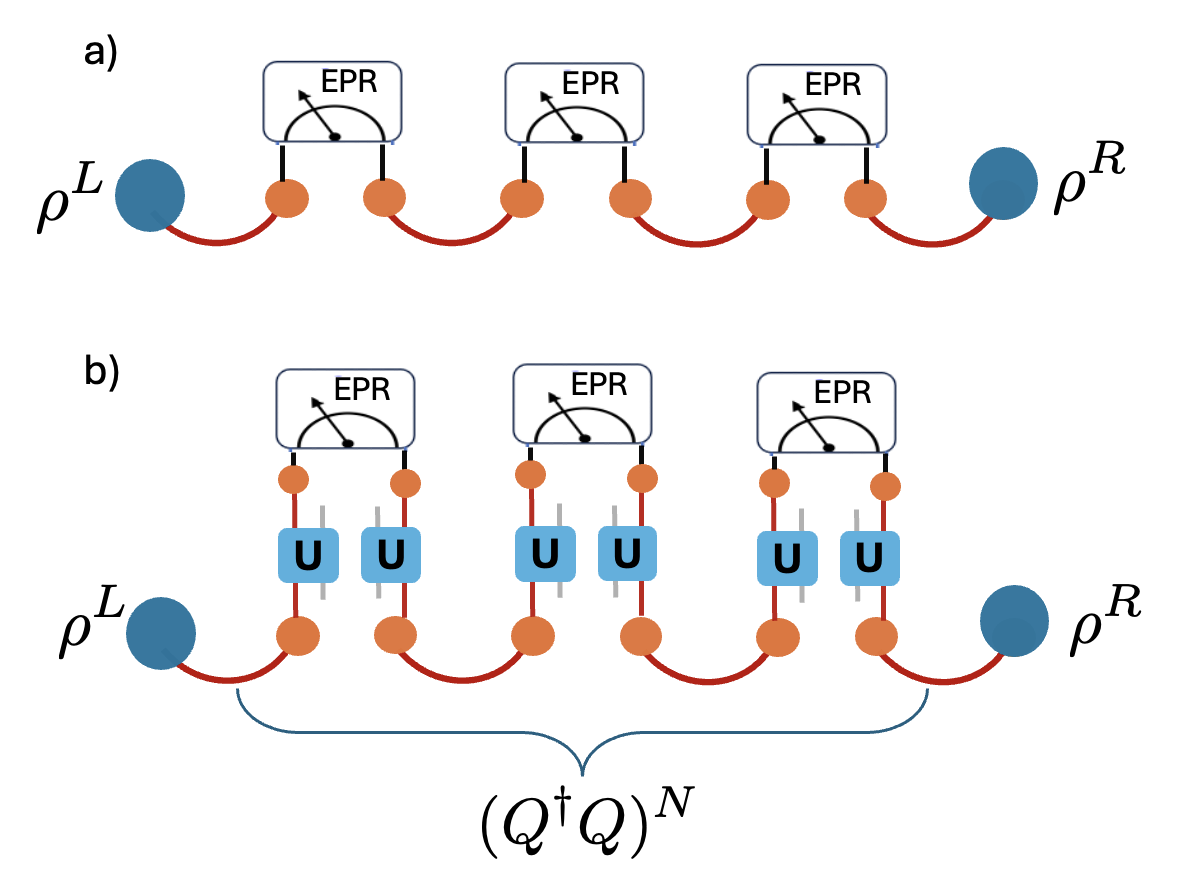}
    \caption{a) An array of EPR pairs connects neighboring even-odd rows. Each orange node represents a row of bits. Projecting the bulk qubits between each odd-even pair onto EPR states effectively generates entanglement between the left and right boundary qubits.
b) Starting from the same array of EPR pairs, we additionally apply an identical quantum channel to each row before performing the same bulk EPR projections. The resulting left and right boundary density matrices, $\rho^L$ and $\rho^R$, are then related by the transfer matrix $\Omega^N$.}
    \label{fig:telep}
\end{figure}

Notably, $\Omega^{N}$ also admits a space-time dual interpretation as the transfer operator governing conditional teleportation. Consider the array shown in Fig.~\ref{fig:telep}-a, where each orange node represents a one-dimensional chain of qubits at row $y$. The rows $y=2i$ and $y=2i+1$ are initially connected site by site by EPR pairs. We then project the complementary pairs of rows, $y=2i-1$ and $y=2i$, onto the same EPR state. Each successful projection performs entanglement swapping: it consumes the neighboring EPR pairs and transfers their entanglement across the projected rows. Repeating this procedure throughout the bulk ultimately produces an EPR pair between the unmeasured degrees of freedom at the left and right boundaries.

We now let every row at $y$ evolve independently under the same 1d quantum channel $Q$ before performing the EPR projections, as shown in Fig.~\ref{fig:telep}-b. As the channel transforms the 2d state into a mixed-state ensemble, the bulk projection leaves two mixed states on the left and right edge rows. These two edge states remain correlated and satisfy:
\begin{align}
    \ket{\rho^{R}}\rangle
    = \Omega^{N}\ket{\rho^{L}}\rangle
    = (Q^{\dagger}Q)^{N}\ket{\rho^{L}}\rangle ,
\end{align}
Thus, $\Omega^{N}$ governs the conditional transmission of quantum information. It encodes the correlations generated between the two edges under bulk EPR projection in the presence of the additional channel $Q$. 

\subsection{Measurement protocol}

\begin{figure}[h]
    \centering
\includegraphics[width=0.5\textwidth]{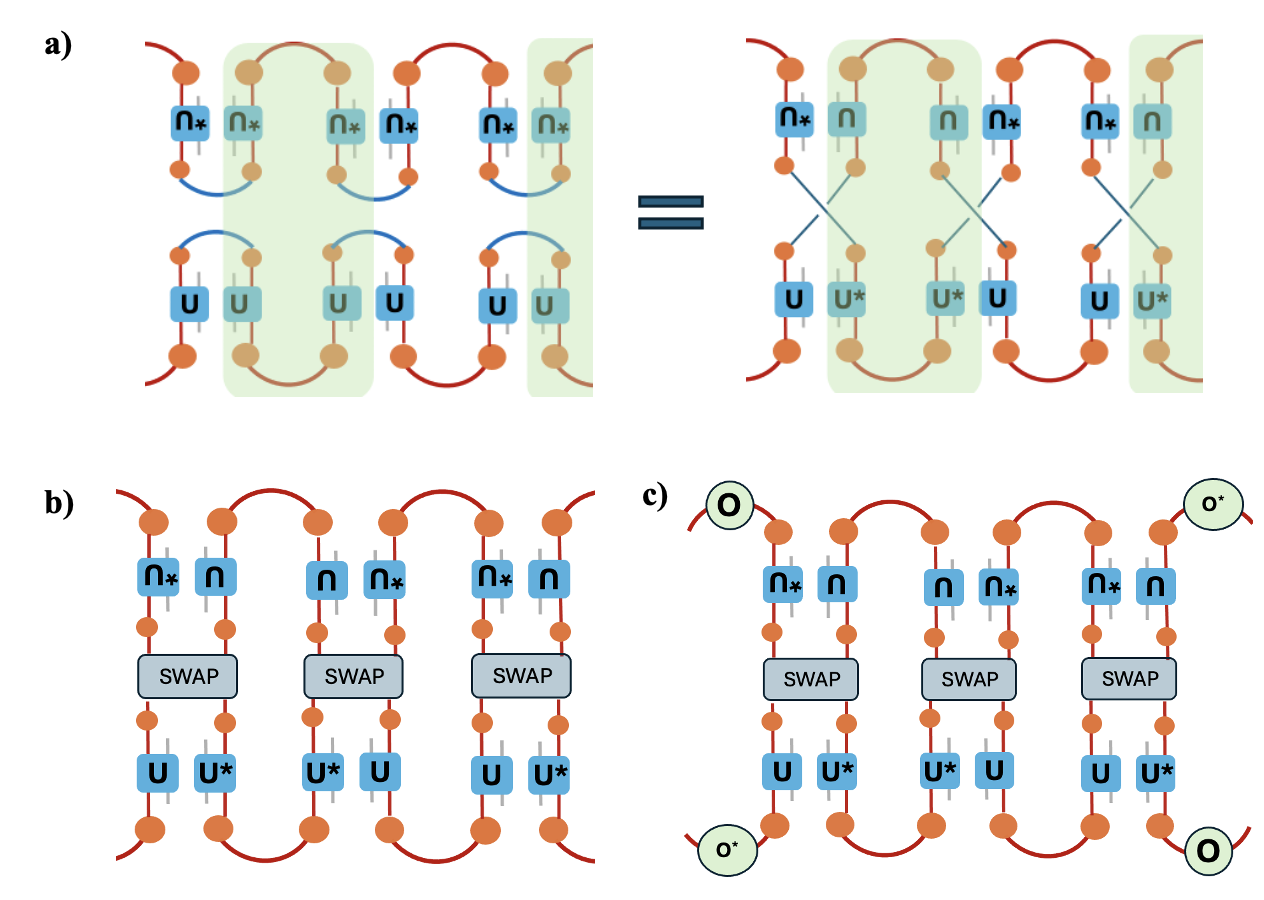}
    \caption{a) The left panel illustrates $\Omega^N$ in the superoperator space, with the top and bottom layers corresponding to the ket and bra spaces, respectively. The right panel shows an equivalent representation obtained by exchanging the ket and bra spaces within the green-shaded region.
b) Multiplication of $\Omega$ is mapped to the measurement of a SWAP operator.
c) The TRNC correlator of $\Omega$ can be measured by appropriately modifying a subset of the initial EPR pairs. }
    \label{fig:swapmeasure}
\end{figure}

The TRNC of $\Omega=Q^{\dagger}Q$, defined in
Eq.~\ref{eq:twsited2}, is challenging to implement experimentally. The
forward map $Q$ is an ordinary quantum channel and can be realized through
a Stinespring dilation by introducing ancilla qubits. In contrast,
$Q^{\dagger}$ is not generally a trace-preserving quantum channel. Its
direct implementation requires measuring the ancillas
with post-selection. Such post-selected operations become increasingly difficult to implement as the number of repetitions grows.

Below we introduce an alternative protocol for measuring the TRNC for $\Omega = Q^{\dagger}Q$, based on the aforementioned spacetime dual picture in Sec.~\ref{sec:epr}. We begin by preparing a spatial array of EPR pairs
oriented along the $y$ direction, such that qubits on rows $y=2i$ and
$y=2i+1$ constitute maximally entangled pairs. The protocol proceeds by
applying the corresponding 1d channel independently to each row,
followed by EPR projections between the intermediate rows $y=2i-1$ and
$y=2i$, as shown in Fig.~\ref{fig:swapmeasure}-a.

As illustrated in Fig.~\ref{fig:swapmeasure}-b, the contractions
generated by these EPR projections can be reproduced by inserting SWAP
operators. In the combined ket and bra representation, the projection
between rows $y=2i-1$ and $y=2i$ pairs the ket index of row $2i-1$ with
the bra index of row $2i$, and vice versa. This crossed pairing is
precisely the index permutation generated by
\begin{align}
    \mathrm{SWAP}_{2i-1,2i}
    =
    \sum_{\alpha,\beta}
    \ket{\alpha}_{2i-1}\ket{\beta}_{2i}
    \bra{\beta}_{2i-1}\bra{\alpha}_{2i}.
\end{align}
Therefore, rather than physically performing the EPR projections and
post-selecting their outcomes, we can obtain the same contraction by
measuring a product of SWAP operators between the corresponding rows.

More explicitly, let
\begin{align}
    \ket{\phi}
    =
    \bigotimes_i
    \ket{\mathrm{EPR}}_{2i,2i+1}
\end{align}
denote the initial product of EPR pairs, and let
\begin{align}
    \rho
    =
    \mathcal{E}\!\left[
        \ket{\phi}\!\bra{\phi}
    \right]
\end{align}
be the state obtained after applying the channel independently to every
row. $\Tr(\Omega^N)$ is then determined by the expectation
value of the staggered SWAP string,
\begin{align}
    \Tr\!\left(\Omega^N\right)
    &\propto
    \Tr\!\left[
        \rho
        \prod_i \mathrm{SWAP}_{2i-1,2i}
    \right].
\end{align}

Likewise, the numerator of the TRNC, which contains operator $O$ insertions
on different replica copies of $\Omega$, can be obtained by modifying
two of the initial EPR pairs into operator-decorated states. For an operator $O$, the required
operator-decorated states are
\begin{align}
    \ket{O}\rangle_{2r,2r+1}
    &=
    \sum_{\alpha,\beta}
    O_{\alpha\beta}
    \ket{\alpha}_{2r}\ket{\beta}_{2r+1},
    \\
    \ket{O^{*}}\rangle_{2r',2r'+1}
    &=
    \sum_{\alpha,\beta}
    O^{*}_{\alpha\beta}
    \ket{\alpha^{*}}_{2r'}\ket{\beta^{*}}_{2r'+1}.
\end{align}
Thus, the EPR pairs on bonds $(2r,2r+1)$ and $(2r',2r'+1)$ are replaced
by $\ket{O}\rangle$ and $\ket{O^{*}}\rangle$, respectively, as
illustrated in Fig.~\ref{fig:swapmeasure}-c. 

While our protocol eliminates the need to implement $Q^{\dagger}$ or perform post-selection, measuring the resulting observable remains challenging at large system sizes. In particular, the protocol requires measuring an extended string of SWAP operators. The resulting observable is therefore highly nonlocal, with a string length that grows with both the number of replicas and the size of each row, making the measurement protocol limited to systems at small scales.

\section{General discussion on LSM in quantum channels}\label{sec:general}

Going beyond the specific example in Sec.~\ref{sec:LSMchain}, our argument suggests a general framework for LSM-type constraints in open quantum systems. We now formulate a generalized Lieb-Schultz-Mattis (LSM) theorem for repeated quantum channels:

\begin{shaded}
Proposition III: \textit{Consider a $d$-dimensional quantum channel with a mixed anomaly between a
\emph{strong} $\mathcal{S}$-symmetry and a \emph{weak} $\mathcal{G}$-symmetry.
Then the Twisted Renyi-$N$ correlator (TRNC) of the channel, defined in
Eq.~\ref{eq:autoco}, exhibits (quasi) long-range order along the
replica direction.}
\end{shaded}

\begin{shaded}
Proposition IV: \textit{For a $d$-dimensional quantum channel with a mixed anomaly between a
\emph{strong} $\mathcal{S}$-symmetry and a \emph{weak} $\mathcal{G}$-symmetry,
the singular-value spectrum of $Q$ (or equivalently, the spectrum of
$\Omega \equiv Q^\dagger Q$) must be either gapless or
degenerate. In addition,
the TRNC defined with respect to $\Omega$ exhibits (quasi) long-range order along the replica direction.}
\end{shaded}

The physical picture behind Propositions III and IV follows from the
spacetime-rotation construction developed in Sec.~\ref{sec:renyi}. Under this
mapping, the repeated evolution of a $d$-dimensional quantum channel with a
mixed anomaly between a strong $\mathcal{S}$-symmetry and a weak
$\mathcal{G}$-symmetry is reinterpreted as a $(d+1)$-dimensional mSPT, protected by the same symmetry combination. This higher-dimensional
mSPT is constructed from the coupled-wire building blocks introduced in
Sec.~\ref{sec:LSMchain}.
Following the argument of Sec.~\ref{sec:Sing}, the singular-value spectrum of the Liouvillian, which is the eigenvalue spectrum of $\Omega \equiv Q^\dagger Q$, is identified with the operator-entanglement spectrum of the density matrix associated with the dual higher-dimensional mSPT.
 The anomaly
therefore enforces this spectrum to be either gapless or degenerate.

The same spacetime-rotation perspective also explains the long-range order
of the TRNC. After the bulk qubits of the $(d+1)$-dimensional mSPT are
projected onto onsite EPR pairs, only the boundary degrees of freedom remain.
These boundary degrees of freedom precisely reproduce the input and output
states of the repeated quantum channel. Thus, the temporal boundary of the
$d$-dimensional channel is mapped to the spatial boundary of the
$(d+1)$-dimensional mSPT after the bulk projection.
From this viewpoint, the quantum channel's TRNC along the temporal direction is dual to the strange correlator of the higher-dimensional mSPT, both of which exhibit LRO.
To make this statement precise, in Appendix~\ref{sec:wzw} we formulate the repeated channel subject to the LSM constraint as a path integral in Choi-double space and show that it maps to an anomalous Wess-Zumino-Witten (WZW) model. Under this correspondence, the TRNC of the
quantum channel becomes the correlation function of the anomalous WZW theory.
Since such correlation
functions cannot be short-ranged, the TRNC must exhibit long-range or
quasi-long-range order~\cite{sala2025entanglement,hsin2024anomalies,zang2024detecting,xu2025diagnosing,zhou2025reviving,kawabata2024lieb,he2020lieb}.

A few clarifications regarding the present LSM theorem are in order:

First, spacetime duality maps the twisted R\'enyi-$N$ correlator (TRNC) of a 1d quantum channel to strange correlators of a 2d SPT state. The SPT nature requires that at least one of the strange correlator, $C^{s}$ in Eq.~\ref{eq:twsited1} or $C^{w}$ in Eq.~\ref{eq:twsited2}, exhibit (quasi)-long-range order. It does not require both correlators to be long-ranged simultaneously. Accordingly, for an LSM-constrained 1d quantum channel, at least one of the two TRNCs, either that associated with the weak $\mathcal{G}$ symmetry or that associated with the strong $\mathcal{S}$ symmetry, must exhibit (quasi)-long-range temporal correlations.

Second, the spectrum of \(Q\) alone is not a distinctive signature of the LSM constraint, since a degenerate spectrum can arise generically in strongly symmetric quantum channels. As we elucidated in Sec.~\ref{sec:Sing}, a sharper LSM diagnostic is the singular-value spectrum of \(Q\), equivalently the spectrum of \(\Omega = Q^\dagger Q\). The TRNC associated with \(\Omega\) must likewise exhibit quasi-long-range or long-range order. 

\section{Conclusion and Outlook}
In this work, we developed a spacetime-duality framework for repeated quantum channels subject to an LSM constraint. The temporal evolution of a 1d channel is reinterpreted as the spatial transfer matrix of a 2d mixed-state SPT (mSPT) generated through bulk projective measurements. Under this correspondence, the temporal boundaries of the channel, namely its input and output states, are mapped to the spatial boundaries of the measured mSPT~\cite{lessa2025mixed,zhang2022strange,ma2022average}, while the TRNC of the 1d channel is mapped to a spatial strange correlator of 2d mSPT. 

Although our construction focuses on the LSM anomaly, the same logic should extend to a broader class of anomalous open systems~\cite{ma2024symmetry,zhou2025reviving,li2025exact,lessa2025mixed,wang2024anomaly,zang2024detecting,panahi2026quantum,lessa2025higher,xu2025average,xu2025diagnosing,lu2025holographic}. For example, intrinsic gapless SPT phases and deconfined quantum critical points, while realizable in UV lattice models, often admit low-energy effective theories with emergent anomalies. 


The spacetime duality introduced here may be viewed as a \textit{dimensional-extension principle}, mapping a \(d\)-dimensional anomalous quantum channel to a \((d+1)\)-dimensional mSPT. This naturally raises the converse question of whether a corresponding \textit{dimensional reduction} is possible. More precisely, given a \((d+1)\)-dimensional mSPT ensemble with weak \(\mathcal{G}\) and strong \(\mathcal{S}\) symmetries, can its spatial strange correlator be identified with the TRNC of a \(d\)-dimensional quantum channel carrying the same symmetry and anomaly structure? In general, the spatial transfer matrix of an mSPT need not correspond to a completely positive and trace-preserving map, and therefore cannot always be interpreted as a physical quantum channel. Nevertheless, for certain fixed-point mSPT density matrices that admit the coupled-wire construction described in Sec.~\ref{sec:LSMchain}, the corresponding transfer matrices take a special form that can be dualized back to lower-dimensional quantum channels, thereby realizing the reverse direction of the duality.

A natural direction for future work is extending our framework to quantum circuits beyond CPTP maps. One compelling setting is a quantum channel supplemented by syndrome measurements, which project the ancilla qubits onto fixed, onsite basis states $\ket{a}_i$ compatible with the weak symmetry $\mathcal{G}$. 
We expect many of the conclusions derived here to remain valid in such non-CPTP settings. Specifically, provided the channel dynamics retain the mixed anomaly between the strong $\mathcal{S}$ and weak $\mathcal{G}$ symmetries, the TRNC should still exhibit long-range order along the replica direction. This expectation follows directly from the spacetime duality argument in Eq.~\ref{eq:twsited1}, which maps the TRNC to the strange correlator of the dual higher-dimensional SPT state. Within this strange-correlator description, projecting the ancilla qubits onto fixed states $\ket{a}_i$ is equivalent to choosing the corresponding product state as the trivial reference state for the ancilla. 
We leave a systematic characterization of the LSM constraint in non-CPTP dynamics for future study.

Finally, our work instantiates a broader program of using spacetime duality to relate well-established equilibrium properties of ground states to the often more challenging nonequilibrium dynamics in one lower dimension. In particular, Ref.~\cite{lu2025holographic} introduced a different spacetime duality relating strongly symmetric quantum-channel dynamics to intrinsic topological order in one higher dimension. Given that SPT and intrinsic topological phases are themselves related by gauging of the symmetries, the two frameworks appear naturally complementary. Understanding the precise connections between them, and ultimately developing a unified dictionary between equilibrium and nonequilibrium physics across dimensions, is an interesting direction for future work.

 \acknowledgments 
This research was supported in part by the National Science Foundation under Grant No. PHY-2309135 through the Kavli Institute for Theoretical Physics (KITP) and was conducted in part at the Aspen Center for Physics, which is supported by National Science Foundation Grant No. PHY-2210452. TCL acknowledges support from the RQS Postdoctoral Fellowship through the National Science Foundation QLCI Grant No. OMA-2120757. YY acknowledges support from the National Science Foundation under Award No. DMR-2439118 and from Cottrell Scholar Award No. CS-CSA-2026-068.  YJL acknowledges support from the MIT Center for Theoretical Physics - a Leinweber Institute, NSF CIQC Award No. 10434 and NSF Early Career DMR-2237244. FP acknowledges support from the Deutsche Forschungsgemeinschaft (DFG, German Research Foundation) under Germany’s Excellence Strategy--EXC--2111--390814868 and the Munich Quantum Valley, which is supported by the Bavarian state government with funds from the Hightech Agenda Bayern Plus.

\appendix

\tikzset{
    coltensor/.style={
        draw,
        rounded corners,
        minimum width=1.0cm,
        minimum height=.60cm,
        inner sep=1pt
    },
    tleg/.style={line width=.45pt}
}

\section{Steady-state structures based on channel symmetries}\label{app:symmetry}

In this section, we present a general discussion on how the symmetry anomaly associated with quantum channels constrains the structure of steady states.

\begin{shaded}
Preposition I: Let $\mathcal{E}$ be a quantum channel with a strong symmetry $S$ and a weak symmetry $W$. Let $\Pi_s$ be a projector into this subspace with a definite strong symmetry charge. If this subspace is invariant under the weak symmetry action $U_w$, namely, $U_w\Pi_s U_w^{\dagger } = \Pi_s$, and if it has a unique steady state $\rho_s$ under $\mathcal{E}$, then $\rho_s$ must have the weak symmetry, regardless of whether the initial state has the weak symmetry or not.
\end{shaded}

{

\itshape

Proof: given $U_w\Pi_s U_w^{\dagger } = \Pi_s$, we can repeatedly apply $\mathcal{E}$: 
\begin{equation}\label{eq:rho_s}
\lim_{N \to \infty}  \mathcal{E}^{N} (U_w\Pi_s U_w^{\dagger } )= \lim_{N \to \infty}  \mathcal{E}^{N}( \Pi_s) =  \rho_s.    
\end{equation}
where the last equality follows from the uniqueness of the steady state $\rho_s$. Now we use the weak symmetry of the channel:
\begin{equation}
\begin{split}
\lim_{N \to \infty}  \mathcal{E}^{N} ( U_w\Pi_s U_w^{\dagger } ) &=    U_w \left(\lim_{N \to \infty}     \mathcal{E}^N (\rho) \right) U_w^{\dagger} \\
&= U_w\rho_s U_w^{\dagger}.
\end{split}
\end{equation}
Combining the two equations above then leads to $ U_w \rho_s  U_w^{\dagger} =    \rho_s$ and finishes the proof. }

Despite having a simple proof, this result is conceptually quite interesting since it illustrates the fact that the weak symmetry of quantum states is a property that can dynamically emerge under the channel. 

Moreover, if the strong $S$ and weak $W$ symmetries together exhibit a mixed anomaly, then the unique steady state $\rho_s$ is expected to have some non-trivial correlation or entanglement structures. This can be seen from the doubled-state representation, in which the pure doubled state $| \rho_s \rangle \rangle$ is symmetric under $S\times S$, and $W$, which together form a mixed anomaly and enforce that $| \rho_s \rangle \rangle$ must be long-range entangled: it cannot be prepared from product states using finite-depth local unitary circuits. In particular, if its long-range entanglement is associated with certain symmetry breaking, then the physical mixed state may have a corresponding symmetry breaking pattern that may give rise to certain long-range (quantum or classical) correlations as well. In our present setting, since by assumption, there is only one unique steady state $\rho_s$ in the strong symmetry sector, we do not expect an order associated with the spontaneous breaking of the weak $W$ symmetry. Instead, we expect that the strong symmetry $S$ of the unique steady state $\rho_s$ is spontaneously broken down to the weak symmetry or is spontaneously broken down to nothing. The former is usually manifested by a long-range conditional mutual information, whereas the latter is usually manifested by a long-range mutual information and correlation in local operators charged under $S$ symmetry. We will discuss one concrete example where the channel has a unique steady state in a strong symmetry sector that exhibits a strong-weak spontaneous symmetry breaking in Sec \ref{sec:num}.

Proposition I also leads to another interesting consequence. 
\begin{shaded}
Corollary I: Using the same notations as in Theorem I, if the subspace obtained by $\Pi_s$ is invariant under the weak symmetry action $U_w$, namely, $U_w\Pi_s U_w^{\dagger } = \Pi_s$, and if there exists a steady state that does not have the weak symmetry, then the channel cannot have a unique steady state in this strong symmetry sector.
\end{shaded}

Notably, the enforced degeneracy in the steady state manifolds does not rely on the mixed anomaly between $S$ and $W$, so it is natural to ask: what are the consequences when imposing a mixed anomaly between the $S$ and $W$ symmetries? We provide a physical perspective below.

First of all, if $\rho$ is a steady state, namely, $\mathcal{E} (\rho) = \rho$, one can apply the conjugation of $U_w$ and find $U_w \mathcal{E} (\rho) U_w^{\dagger} = U_w\rho U_w^{\dagger}  $. Using the weak symmetry property of the channel, one then has  $\mathcal{E} ( U_w  \rho U_w^{\dagger} )  =  U_w\rho U_w^{\dagger} ~~ \forall w \in W$. In other words, the weak symmetry transformation can preserve the steady-state manifold. By assumption, there is no unique steady state, and in this case, one natural possibility is that there are $|W|$ distinct steady states $\{ U_w \rho  U_w^{\dagger}  | w  \in W  \} $ generated by the weak symmetry action, for a suitable $\rho$. In other words, the steady-state degeneracy arises from the spontaneous breaking of the weak $W$ symmetry. Given this, the uniform mixture over all the weak symmetry actions defines a valid steady state $ \rho_W= \frac{1}{|W|} \sum_{w\in W}   U_w \rho  U_w^{\dagger}$, which notably, restores the weak symmetry, i.e. $ U_w \rho_W U_w^{\dagger}  =\rho_W~~  \forall  w  \in W$. Since the above discussion is restricted to a strong symmetry sector, $\rho_W$ is guaranteed to exhibit both the strong $S$ symmetry and the weak $W$ symmetry, thereby indicating certain non-trivial correlation structure due to the symmetry anomaly. A concrete example where the weak symmetry is spontaneously broken is presented in Appendix \ref{app:weakbreak}.



\section{WZW theory for LSM under thermal equilibrium} \label{app:wzwintro}
 
We recall the correspondence between Lieb-Schultz-Mattis (LSM) constraints and Wess-Zumino-Witten (WZW) theory in thermal equilibrium. As demonstrated in Ref.~\cite{jian2018lieb,metlitski2018intrinsic,yang2018dyonic}, any system subject to LSM constraints necessarily carries an emergent ’t~Hooft anomaly. For the concrete case of a spin-$1/2$ chain, this anomaly is captured by an $O(4)$ rotor field $\vec m$ governed by a non-linear sigma model with Wess-Zumino-Witten terms in $1{+}1$d:
\begin{align}
&\mathcal{L}_{WZW}=\frac{1}{g}(\partial_{\mu} \vec{m})^2+\frac{2\pi}{\Omega^4} \int_0^1 du\epsilon^{ijkl}  m_{i}\partial_x m_{j} \partial_t m_{k}\partial_u m_{l},\nonumber\\
&\bm{m}(u=0,x,t)=(0,0,0,1),~ \bm{m}(u=1,x,t)=\bm{m}(x,t)
\label{eq:NLSMwzw} 
\end{align}
Here, the first three components of the $O(4)$ rotor $(m_1,m_2,m_3)$ denote the spin polarizations $(S_x,S_y,(-1)^x S_z)$\footnote{Here, $m_3 = (-1)^x S_z$ has a staggered sign structure between even and odd sites. This staggered structure originates from the antiferromagnetically averaged background with zero net magnetization, so $m_3$ represents the smooth spatial variation of the magnetization. Thus, $m_3$ is staggered and transforms oddly under lattice translation $T_x$.} and transform as an $\mathrm{SO}(3)$ vector, while $m_4$ can be viewed as the valence-bond–solid order parameter that is odd under lattice translations. 
As demonstrated in Ref.~\cite{jian2018lieb,you2014wave,bi2015classification}, the WZW theory in Eq.~\ref{eq:NLSMwzw} can be viewed as a boundary theory that describes the 1d edge of a 2d symmetry protected topological phase (SPT) phase protected by $SO_3 \times \mathbb{Z}_2$. The $\mathbb{Z}_2$ symmetry descends from lattice translations $T_x$ that exchange even and odd sites along the $x$ axis; under renormalization it becomes an internal $\mathbb{Z}_2$ symmetry in the effective theory \cite{thorngren2016gauging}. Euclidean path integral arguments \cite{xu2013wave,jian2018lieb,metlitski2018intrinsic,yang2018dyonic} imply that any $d$ dimensional system obeying LSM constraints can be treated as the boundary of a $d{+}1$ dimensional SPT phase with an emergent anomaly. The gapless or degenerate energy spectrum required by LSM in thermal equilibrium then follows directly from this boundary anomaly.

\section{Tensor-network representations of the coupled-wire construction}
\label{app:tensor_network}

This appendix reformulates the coupled-wire construction of
Sec.~\ref{sec;couplewire} in a tensor-network language. The elementary
building block of the SPT state is associated with a single
Stinespring-dilated channel update, given in Eq.~\ref{eq:block}.
Stacking these building blocks produces the pure two-dimensional SPT
wavefunction of Eq.~\ref{eq:Phinew}. Both the SPT wavefunctions and the corresponding quantum channel can be expressed in
terms of MPS tensors.

\paragraph{2d SPT wavefunction via MPS tensor:}

We regard the coupled-wire wavefunction  in Eq.~\ref{eq:Phinew} as an MPS along the \(y\) direction.
Each MPS site is an entire one-dimensional column at \(y=i\), rather than a
single microscopic spin. Its physical Hilbert space is
\begin{align}
\mathcal H_i^{\mathrm{phys}}
=
\mathcal H_{L_i}\otimes
\mathcal H_{R_i}\otimes
\mathcal H_{A_i}.
\label{eq:column_hilbert}
\end{align}
Thus, every column tensor has three physical 1d wire legs:
the left component \(L_i\), the right left component \(R_i\), and the ancilla wire \(A_i\). Its virtual legs carry the Hilbert space
\(\mathcal H_S\) of an entire spin-\(\frac12\) system chain.

For each step of quantum channel, let \(\bm s\) denote an input system configuration and let \(\bm 0\) denote
the fixed input ancilla configuration. The output system and output ancilla
configurations are denoted by \(\bm R_i\) and
\(A_i\), respectively. The unitary gate that generates the quantum channel in the Stinespring-dilated picture is:
\begin{align}
\mathcal U^{
\,\bm R_i,A_i
}_{\bm s,\bm 0}
\equiv
\langle
\bm R_i,A_i
|
\hat U^{s,a}
|
\bm s,\bm 0
\rangle .
\label{eq:stinespring_tensor}
\end{align}
The upper indices of \(\mathcal U\) therefore label the system and ancilla
outputs, while the lower indices label the system and ancilla inputs.

The corresponding MPS tensor for the dual wavefunction is:
\begin{align}
\left(\mathcal A_i\right)^{
\bm L_i,\bm R_i,A_i
}_{\bm\mu_{i-1},\bm\mu_i}
=
\delta_{\bm L_i,\bm\mu_i}\,
\mathcal U^{
\,\bm R_i,A_i
}_{\bm\mu_{i-1},\bm 0}.
\label{eq:column_MPS_tensor}
\end{align}

\begin{equation}
\vcenter{\hbox{
\begin{tikzpicture}[x=1cm,y=1cm,baseline=(current bounding box.center)]

\node[coltensor] (Ai) at (-2.2,-.15) {\(\mathcal A_i\)};

\draw[tleg] (Ai.west)--++(-.75,0);
\draw[tleg] (Ai.east)--++(.75,0);

\node[font=\scriptsize] at (-3.15,.08) {\(\nu_i\)};
\node[font=\scriptsize] at (-1.35,.08) {\(\mu_i\)};

\draw[tleg] (-2.50,.15)--(-2.50,.90);
\draw[tleg] (-2.20,.15)--(-2.20,.90);
\draw[tleg] (-1.90,.15)--(-1.90,.90);

\node[font=\scriptsize] at (-2.50,1.08) {\(L_i\)};
\node[font=\scriptsize] at (-2.20,1.08) {\(A_i\)};
\node[font=\scriptsize] at (-1.90,1.08) {\(R_i\)};

\node at (-.40,.20) {\(\displaystyle =\)};


\draw[tleg]
(1.20,.90)--(1.20,-.12)
.. controls (1.20,-.50) and (.82,-.50) ..
(.62,-.15)--(.25,-.15);

\draw[tleg] (1.75,.90)--(1.75,-.54);
\fill (1.75,-.54) circle (.075);

\draw[tleg]
(2.30,.90)--(2.30,-.12)
.. controls (2.30,-.50) and (2.68,-.50) ..
(2.88,-.15)--(3.25,-.15);

\node[font=\scriptsize] at (.43,.08) {\(\nu_i\)};
\node[font=\scriptsize] at (3.07,.08) {\(\mu_i\)};

\node[
    draw=blue,
    fill=cyan!65,
    text=black,
    minimum width=1.05cm,
    minimum height=.55cm,
    font=\large
] at (2.025,.40) {\(\hat U^{s,a}\)};

\node[font=\scriptsize] at (1.20,1.08) {\(L_i\)};
\node[font=\scriptsize] at (1.75,1.08) {\(A_i\)};
\node[font=\scriptsize] at (2.30,1.08) {\(R_i\)};

\end{tikzpicture}
}}
\label{eq:column_tensor_diagram}
\end{equation}

The horizontal legs in Eq.~\ref{eq:column_tensor_diagram} are virtual
system-wire legs. The three vertical legs are physical wires. This distinction
is essential below: virtual-leg contraction generates the coupled-wire
wavefunction, whereas physical-leg contraction generates repeated channel
evolution.

\paragraph{Pure-state SPT:}

Contracting only the virtual indices gives the 2d SPT state as Eq.~\ref{eq:Phinew} obtained from coupled wire constructions:
\begin{align}
|\Phi\rangle
=
\operatorname{tTr}_{\{\bm\mu_i\}}
\left[
\bigotimes_i\mathcal A_i
\right].
\label{eq:pure_SPT_tensor}
\end{align}
Here \(\operatorname{tTr}\) contracts only the virtual wire indices. All
physical \(L_i\), \(R_i\), and \(A_i\) legs remain open.
\begin{equation}
|\Phi\rangle =
\vcenter{\hbox{
\begin{tikzpicture}[x=1cm,y=1cm,baseline=(current bounding box.center)]
\node at (-2.35,0) {\(\cdots\)};
\node[coltensor] (A0) at (-1.35,0) {\(\mathcal A_{i-1}\)};
\node[coltensor] (A1) at (0,0) {\(\mathcal A_i\)};
\node[coltensor] (A2) at (1.35,0) {\(\mathcal A_{i+1}\)};
\node at (2.35,0) {\(\cdots\)};

\draw[tleg] (A0.east)--(A1.west);
\draw[tleg] (A1.east)--(A2.west);
\node[font=\scriptsize] at (.68,.22) {\(\mu_i\)};

\draw[tleg] (-1.69,-.30)--(-1.69,-.78);
\draw[tleg] (-1.35,-.30)--(-1.35,-.78);
\draw[tleg] (-1.01,-.30)--(-1.01,-.78);

\draw[tleg] (-.34,-.30)--(-.34,-.78);
\draw[tleg] (0,-.30)--(0,-.78);
\draw[tleg] (.34,-.30)--(.34,-.78);
\node[font=\scriptsize] at (-.34,-.98) {\(L_i\)};
\node[font=\scriptsize] at (0,-.98) {\(R_i\)};
\node[font=\scriptsize] at (.34,-.98) {\(A_i\)};

\draw[tleg] (1.01,-.30)--(1.01,-.78);
\draw[tleg] (1.35,-.30)--(1.35,-.78);
\draw[tleg] (1.69,-.30)--(1.69,-.78);
\end{tikzpicture}
}}
\label{eq:pure_SPT_diagram}
\end{equation}

Eq.~\ref{eq:pure_SPT_diagram} gives the tensor-network representation of Eq.~\ref{eq:Phinew}. The virtual wires connecting neighboring tensors encode the coupled-wire structure. In the absence of ancilla and when the quantum channel is the identity map, the tensor reduces to:
\begin{align}
\left(\mathcal A_i\right)^{
\bm L_i,\bm R_i
}_{\bm\mu_{i-1},\bm\mu_i}
=
\delta_{\bm L_i,\bm\mu_i}
\delta_{\bm R_i,\bm\mu_{i-1}},
\label{eq:identity_column_tensor}
\end{align}
 This is exactly the EPR pairing between
\(R_{i+1}\) and \(L_i\) in Eq.~\ref{eq:Phi} and
Fig.~\ref{dualmain2}-b. 

\paragraph{Mixed-state SPT.}

The mSPT in Eq.~\ref{eq:mSPT} is obtained by tracing the \emph{physical} ancilla legs between the
ket and bra tensor networks:
\begin{align}
\rho^{\mathrm{mSPT}}
=
\Tr_{\{A_i\}}
\left[
|\Phi\rangle\langle\Phi|
\right].
\label{eq:mSPT_tensor}
\end{align}
In the MPO formalism, we contract the ancilla legs while leaving the
$L_i$ and $R_i$ legs open. These open legs carry the density-matrix
indices associated with both the ket and bra spaces.

\begin{equation}
\rho^{\mathrm{mSPT}}=
\vcenter{\hbox{
\begin{tikzpicture}[x=1cm,y=1cm,baseline=(current bounding box.center)]

\node[coltensor] (Km) at (-1.6,-.60) {\(\mathcal A_{i-1}\)};
\node[coltensor] (K)  at (0,-.60)    {\(\mathcal A_i\)};
\node[coltensor] (Kp) at (1.6,-.60)  {\(\mathcal A_{i+1}\)};

\node[coltensor] (Bm) at (-1.6,.60) {\(\mathcal A_{i-1}^{\dagger}\)};
\node[coltensor] (B)  at (0,.60)    {\(\mathcal A_i^{\dagger}\)};
\node[coltensor] (Bp) at (1.6,.60)  {\(\mathcal A_{i+1}^{\dagger}\)};

\draw[tleg] (Km.east)--(K.west);
\draw[tleg] (K.east)--(Kp.west);
\draw[tleg] (Bm.east)--(B.west);
\draw[tleg] (B.east)--(Bp.west);

\draw[tleg] (Km.west)--++(-.38,0);
\draw[tleg] (Bm.west)--++(-.38,0);
\draw[tleg] (Kp.east)--++(.38,0);
\draw[tleg] (Bp.east)--++(.38,0);

\draw[tleg] (Km.north east)--(Bm.south east);
\draw[tleg] (K.north east)--(B.south east);
\draw[tleg] (Kp.north east)--(Bp.south east);
\node[font=\scriptsize,right] at (.58,0) {\(\Tr_{A_i}\)};

\draw[tleg] (Km.235)--++(0,-.34);
\draw[tleg] (Km.265)--++(0,-.34);
\draw[tleg] (Bm.125)--++(0,.34);
\draw[tleg] (Bm.95)--++(0,.34);

\draw[tleg] (K.235)--++(0,-.34);
\draw[tleg] (K.265)--++(0,-.34);
\draw[tleg] (B.125)--++(0,.34);
\draw[tleg] (B.95)--++(0,.34);

\draw[tleg] (Kp.235)--++(0,-.34);
\draw[tleg] (Kp.265)--++(0,-.34);
\draw[tleg] (Bp.125)--++(0,.34);
\draw[tleg] (Bp.95)--++(0,.34);

\node[font=\scriptsize] at (-.14,-1.50) {\(L_i,\ R_i\)};

\end{tikzpicture}
}}
\label{eq:mSPT_diagram}
\end{equation}
Eq~\ref{eq:mSPT_diagram} gives the diagrammatic representation of each MPO tensor for the mSPT state. Tracing out the ancilla degrees of freedom converts the pure SPT state into a mixed state, such that each MPO tensor carries both ket and bra indices.

\paragraph{Repeated channel from EPR projection:}
To relate the repeated application of the quantum channel to the SPT state, we impose an onsite EPR projection. We define the onsite EPR state as
\begin{align}
|\Omega\rangle_{L_iR_i}
\propto
\sum_{\bm s}
|\bm s\rangle_{L_i}|\bm s\rangle_{R_i},
\qquad
\mathcal P_\Omega
=
\prod_{i=1}^{N-1}
{}_{L_iR_i}\langle\Omega|.
\label{eq:Bell_projection}
\end{align}
The operator \(\mathcal P_\Omega\) contracts the \emph{physical} \(L_i\) and
\(R_i\) legs within the same column in the bulk. Since \(R_i\) is the output-system leg
of the \(i\)-th channel block and \(L_i\) is the input leg of the
\((i+1)\)-th block, this physical EPR contraction concatenates adjacent
channel updates as:
\begin{equation}
\vcenter{\hbox{
\begin{tikzpicture}[x=1cm,y=1cm,baseline=(current bounding box.center)]
\node at (-2.15,0) {\(\cdots\)};
\node[coltensor] (Am) at (-1.30,0) {\(\mathcal A_{i-1}\)};
\node[coltensor] (A)  at (0,0)     {\(\mathcal A_i\)};
\node[coltensor] (Ap) at (1.30,0)  {\(\mathcal A_{i+1}\)};
\node at (2.15,0) {\(\cdots\)};

\draw[tleg] (Am.west) -- ++(-.40,0);
\draw[tleg] (Am.east) -- (A.west);
\draw[tleg] (A.east)  -- (Ap.west);
\draw[tleg] (Ap.east) -- ++(.40,0);

\path (Am.235) ++(0,-.32) coordinate (AmL);
\path (Am.265) ++(0,-.32) coordinate (AmR);
\draw[tleg] (Am.235) -- (AmL)
    .. controls +(0,-.20) and +(0,-.20) .. (AmR)
    -- (Am.265);
\draw[tleg] (Am.305) -- ++(0,-.50) coordinate (AmA);
\node[font=\scriptsize,below] at (AmA) {\(A_{i-1}\)};

\path (A.235) ++(0,-.32) coordinate (AL);
\path (A.265) ++(0,-.32) coordinate (AR);
\draw[tleg] (A.235) -- (AL)
    .. controls +(0,-.20) and +(0,-.20) .. (AR)
    -- (A.265);
\draw[tleg] (A.305) -- ++(0,-.50) coordinate (AA);
\node[font=\scriptsize,below] at (AA) {\(A_i\)};

\path (Ap.235) ++(0,-.32) coordinate (ApL);
\path (Ap.265) ++(0,-.32) coordinate (ApR);
\draw[tleg] (Ap.235) -- (ApL)
    .. controls +(0,-.20) and +(0,-.20) .. (ApR)
    -- (Ap.265);
\draw[tleg] (Ap.305) -- ++(0,-.50) coordinate (ApA);
\node[font=\scriptsize,below] at (ApA) {\(A_{i+1}\)};
\end{tikzpicture}
}}
\label{eq:open_channel_diagram}
\end{equation}
The EPR projections, together with tracing over the ancilla legs, reproduce the repeated application of the quantum channel in the Choi-doubled space. This process can be represented diagrammatically as
\begin{equation}
Q^N=
\vcenter{\hbox{
\begin{tikzpicture}[x=1cm,y=1cm,baseline=(current bounding box.center)]
\node[coltensor] (Km) at (-1.55,-.60) {\(\mathcal A_{i-1}\)};
\node[coltensor] (K)  at (0,-.60)     {\(\mathcal A_i\)};
\node[coltensor] (Kp) at (1.55,-.60)  {\(\mathcal A_{i+1}\)};

\node[coltensor] (Bm) at (-1.55,.60) {\(\mathcal A_{i-1}^\dagger\)};
\node[coltensor] (B)  at (0,.60)     {\(\mathcal A_i^\dagger\)};
\node[coltensor] (Bp) at (1.55,.60)  {\(\mathcal A_{i+1}^\dagger\)};

\node at (-2.45,-.60) {\(\cdots\)};
\node at (2.45,-.60)  {\(\cdots\)};
\node at (-2.45,.60)  {\(\cdots\)};
\node at (2.45,.60)   {\(\cdots\)};

\draw[tleg] (Km.west) -- ++(-.35,0);
\draw[tleg] (Km.east) -- (K.west);
\draw[tleg] (K.east)  -- (Kp.west);
\draw[tleg] (Kp.east) -- ++(.35,0);

\draw[tleg] (Bm.west) -- ++(-.35,0);
\draw[tleg] (Bm.east) -- (B.west);
\draw[tleg] (B.east)  -- (Bp.west);
\draw[tleg] (Bp.east) -- ++(.35,0);

\draw[tleg] (Km.55) -- (Bm.305);
\draw[tleg] (K.55)  -- (B.305);
\draw[tleg] (Kp.55) -- (Bp.305);
\path (K.55) -- (B.305) coordinate[pos=.5] (Atr);
\node[font=\scriptsize,right] at (Atr) {\(\Tr_A\)};

\path (Km.235) ++(0,-.28) coordinate (KmL);
\path (Km.265) ++(0,-.28) coordinate (KmR);
\draw[tleg] (Km.235) -- (KmL)
    .. controls +(0,-.18) and +(0,-.18) .. (KmR)
    -- (Km.265);

\path (K.235) ++(0,-.28) coordinate (KL);
\path (K.265) ++(0,-.28) coordinate (KR);
\draw[tleg] (K.235) -- (KL)
    .. controls +(0,-.18) and +(0,-.18) .. (KR)
    -- (K.265);

\path (Kp.235) ++(0,-.28) coordinate (KpL);
\path (Kp.265) ++(0,-.28) coordinate (KpR);
\draw[tleg] (Kp.235) -- (KpL)
    .. controls +(0,-.18) and +(0,-.18) .. (KpR)
    -- (Kp.265);

\path (Bm.125) ++(0,.28) coordinate (BmL);
\path (Bm.95)  ++(0,.28) coordinate (BmR);
\draw[tleg] (Bm.125) -- (BmL)
    .. controls +(0,.18) and +(0,.18) .. (BmR)
    -- (Bm.95);

\path (B.125) ++(0,.28) coordinate (BL);
\path (B.95)  ++(0,.28) coordinate (BR);
\draw[tleg] (B.125) -- (BL)
    .. controls +(0,.18) and +(0,.18) .. (BR)
    -- (B.95);

\path (Bp.125) ++(0,.28) coordinate (BpL);
\path (Bp.95)  ++(0,.28) coordinate (BpR);
\draw[tleg] (Bp.125) -- (BpL)
    .. controls +(0,.18) and +(0,.18) .. (BpR)
    -- (Bp.95);
\end{tikzpicture}
}}
\label{eq:standard_channel_diagram}
\end{equation}
Eq.~\ref{eq:standard_channel_diagram} thus provides the tensor representation of the transfer matrix $Q^N$ associated with the quantum channel.

\section{Path integral of quantum channel from WZW theory}
\label{sec:wzw}

We now reformulate the lattice coupled-wire construction in Sec.~\ref{sec;couplewire} in terms of the continuum WZW theory. In the absence of ancilla degrees of freedom, one elementary building block in Eq.~\ref{eq:Phi} consists of two coupled spin-$1/2$ wires, $R_i$ and $L_{i-1}$. Replacing each wire by its WZW description, the wavefunction of this building block takes the form
\begin{align}
&|A\rangle_{R_i,L_{i-1}}
=
\sum_{\bm m(x)}
|\bm m(x)\rangle_{R_i}
|\bm m(x)\rangle_{L_{i-1}}
\nonumber\\
&\sim
\int
\mathcal D[\bm m^{i-1}_L,\bm m^{i}_R]\,
e^{-\mathcal S_{i,i-1}[\bm m^{i-1}_L,\bm m^{i}_R]}
|\bm m^{i}_R\rangle
|\bm m^{i-1}_L\rangle .
\label{eq:rho_WZW}
\end{align}
The first line is the continuum representation of the EPR pairing in Eq.~\ref{eq:Phi}. The second line expresses its wavefunction as a path integral over the field configurations $\bm m_L$ and $\bm m_R$ on the two wires, with
\begin{align}
\mathcal S_{i,i-1}[\bm m^{i-1}_L,\bm m^{i}_R]
=
\mathcal S[\bm m^{i-1}_L]
+
\mathcal S^{*}[\bm m^{i}_R]
-
U\,\bm m^{i-1}_L\cdot\bm m^{i}_R .
\label{choidouble}
\end{align}
Here, $\mathcal S[\bm m^{i-1}_L]$ and $\mathcal S^{*}[\bm m^{i}_R]$ are the effective WZW actions in Eq.~\ref{eq:NLSMwzw} for the two spin-$1/2$ wires in the building block. The complex conjugation reflects the conjugate WZW theories carried by the $L$ and $R$ components. The phenomenological coupling $U$ entangles the two wires. In the fixed-point limit $U\rightarrow\infty$, it locks $\bm m^{i-1}_L$ and $\bm m^{i}_R$, reproducing the EPR state in Eq.~\ref{eq:Phi}.

More generally, a local symmetry-preserving unitary may act within one spin chain during each channel step, deforming the fixed-point wavefunction into
\begin{align}
\sum_{\bm m(x)}
\left(
\hat{\mathbb U}
|\bm m(x)\rangle_{R_i}
\right)
\otimes
|\bm m(x)\rangle_{L_{i-1}} .
\end{align}
The resulting state in the building block need not remain a product of EPR pairs between adjacent columns. Accordingly, the coupling $U$ in Eq.~\ref{choidouble} can be kept finite to phenomenologically describe such local deformations of the fixed-point state.

In the coupled-wire construction, each column at $y=i$ contains two one-dimensional theories, labeled by $L_i$ and $R_i$. The elementary building block couples $R_i$ to $L_{i-1}$, forming a coupled WZW theory on a two-leg ladder, as described in Eq.~\ref{choidouble}. To obtain a continuum description in the transverse $y$-direction, we promote the discrete interval between the two wires to a \textit{continuous strip} $y\in[i-1,i]$. In this continuum description, the two coupled wires on the ladder are identified with the spatial boundary fields at $y=i-1$ and $y=i$, respectively, subject to the boundary conditions
\begin{align}
\bm m^{i-1}_L(x)=\bm m(x,i-1),
\qquad
\bm m^{i}_R(x)=\bm m(x,i).
\end{align}
The action of one building block in Eq.~\ref{choidouble} can then be approximated as a continuum field $\bm{m}(x,y)$ on the stripe:
\begin{align}
\mathcal S_{i,i-1}[\bm m]
&\sim
\int_{i-1}^{i} dy\int dx\,
\bigg[
\frac{1}{g}(\partial_x\bm m)^2
+
\frac{1}{g}(\partial_y\bm m)^2
\nonumber\\
&+
\frac{2\pi i}{\Omega^4}
\int_{0}^{1}du\,
\epsilon^{abcd}
m_a\partial_xm_b\partial_um_c\partial_ym_d
\bigg].
\label{eq:coupled}
\end{align}
In this continuous extension, the boundary coupling
$U\,\bm m(x,i-1)\cdot\bm m(x,i)$ is promoted phenomenologically to the local bulk term $(\partial_y\bm m)^2$. The first two terms in Eq.~\ref{eq:coupled} are therefore the gradient terms of the two-dimensional nonlinear sigma model, while the last term is the $\mathrm{O}(4)$ WZW term defined on the extended $(x,y,u)$ space. Together, they describe the WZW theory on a finite strip, with $1/g\sim U$.

We next introduce a quantum channel that decoheres the pure state of each building block into a mixed state. The density matrix of one building block is
\begin{widetext}
\begin{align}
\rho^{A_{R_i,L_{i-1}}}
\sim
\int
\mathcal D[
\bm m, \bm n]\,
&e^{-\mathcal S_{i,i-1}[\bm m]
-\mathcal S_{i,i-1}[\bm n]
-\mathcal S_{\mathrm{int}}[\bm n,\bm m]}
~
|\bm m^{i}_R\rangle
|\bm m^{i-1}_L\rangle
\langle\bm n^{i}_R|
\langle\bm n^{i-1}_L|,
\nonumber\\
\mathcal S_{\mathrm{int}}[\bm n,\bm m]
&\sim
\int_{i-1}^{i}dy\int dx\,
g'\,m_4n_4 .
\label{eq:rho_WZW_mixed}
\end{align}
\end{widetext}
$\mathcal S_{\mathrm{int}}[\bm n,\bm m]$ denotes the coupling between the bra and ket fields induced by decoherence.
Here, $\bm m$ and $\bm n$ denote the continuum ket and bra fields, respectively, defined on the segment $y\in[i-1,i]$. The original wire degrees of freedom are identified with the boundary values of these continuum fields. Explicitly,
\begin{align}
\bm m^{i-1}_L(x) = \bm m(x,i-1),~
\bm m^{i}_R(x) = \bm m(x,i),\nonumber\\
\bm n^{i-1}_L(x) = \bm n(x,i-1),~
\bm n^{i}_R(x) = \bm n(x,i).
\end{align}
Thus, the fields on the $L_{i-1}$ and $R_i$ wires serve as the spatial boundary conditions for the continuum extension along the $y$ direction.

The symmetry transformations are
\begin{align}
T_x:\qquad
(n_1,n_2,n_3,n_4)
&\rightarrow
(n_1,n_2,-n_3,-n_4),
\nonumber\\
(m_1,m_2,m_3,m_4)
&\rightarrow
(m_1,m_2,-m_3,-m_4),
\nonumber\\
\mathrm{SO}(3)_{\mathrm{ket}}:\qquad
(m_1,m_2,m_3)^T
&\rightarrow
R(\theta,\phi)(m_1,m_2,m_3)^T,
\nonumber\\
\mathrm{SO}(3)_{\mathrm{bra}}:\qquad
(n_1,n_2,n_3)^T
&\rightarrow
R(\theta',\phi')(n_1,n_2,n_3)^T .
\label{eq:WZW_symmetry}
\end{align}
The weak translation symmetry $T_x$ acts simultaneously on the ket and bra fields, whereas the strong $\mathrm{SO}(3)$ symmetry acts independently in the two sectors. Since $m_4$ and $n_4$ are neutral under $\mathrm{SO}(3)$, the coupling $m_4n_4$ is compatible with strong $\mathrm{SO}(3)$. At the same time, it is invariant under the simultaneous action of $T_x$ on the ket and bra fields and therefore preserves weak translation symmetry.

After implementing the spacetime duality introduced in Sec.~\ref{sec;couplewire}, the density matrix of one building block can be reorganized as the transfer matrix $Q$ of the single step of a quantum channel defined in the super operator space:
\begin{widetext}
\begin{align}
Q
\sim
\int
\mathcal D[
\bm m, \bm n]\,
&e^{-\mathcal S_{i,i-1}[\bm m]
-\mathcal S_{i,i-1}[\bm n]
-\mathcal S_{\mathrm{int}}[\bm n,\bm m]}
~~
|\bm n^i_R\rangle
|\bm m^i_R\rangle
\langle\bm n^{i-1}_L|
\langle\bm m^{i-1}_L|.
\label{eq:Q_WZW}
\end{align}
\end{widetext}
The fields $\bm m$ and $\bm n$ form the two layers of the Choi-double space, originating from the ket and bra sectors. Under the spacetime rotation, the spatial boundaries $y=i-1$ and $y=i$ become the time slices immediately before and after one application of the channel.

Repeated application of the channel glues adjacent building blocks by identifying their shared boundary configurations. The transfer matrix connecting the initial and final states after $N$ channel steps is therefore
\begin{widetext}
\begin{align}
Q^N
&\sim
\int
\mathcal D[\bm m,\bm n]\,
e^{-\mathcal S[\bm m]
-\mathcal S[\bm n]
-\mathcal S_{\mathrm{int}}[\bm n,\bm m]}
|\bm n_R\rangle
|\bm m_R\rangle
\langle\bm n_L|
\langle\bm m_L|,
\nonumber\\
\bm m_L
&=
\bm m(x,0),
\qquad
\bm m_R
=
\bm m(x,N),
\nonumber\\
\bm n_L
&=
\bm n(x,0),
\qquad
\bm n_R
=
\bm n(x,N),
\nonumber\\
\mathcal S[\bm m]
&\sim
\int_0^Ndy\int dx\,
\bigg[
\frac{1}{g}(\partial_x\bm m)^2
+
\frac{1}{g}(\partial_y\bm m)^2
\nonumber\\
&\hspace{3.2cm}
+
\frac{2\pi i}{\Omega^4}
\int_0^1du\,
\epsilon^{abcd}
m_a\partial_xm_b\partial_um_c\partial_ym_d
\bigg],
\nonumber\\
\mathcal S[\bm n]
&\sim
\int_0^Ndy\int dx\,
\bigg[
\frac{1}{g}(\partial_x\bm n)^2
+
\frac{1}{g}(\partial_y\bm n)^2
\nonumber\\
&\hspace{3.2cm}
+
\frac{2\pi i}{\Omega^4}
\int_0^1du\,
\epsilon^{abcd}
n_a\partial_xn_b\partial_un_c\partial_yn_d
\bigg],
\nonumber\\
\mathcal S_{\mathrm{int}}[\bm n,\bm m]
&\sim
\int_0^Ndy\int dx\,
g'\,m_4n_4 .
\label{1deft}
\end{align}
\end{widetext}
Eq.~\ref{1deft} shows that repeated application of the quantum channel is represented by the path integral of the coupled two-dimensional WZW theory on a strip of width $N$ in the Choi-double space. The spatial boundaries of the WZW theory at $y=0$ and $y=N$ are mapped to the temporal boundaries of the repeated channel at $t=0$ and $t=N$, respectively. 

The $\mathrm{O}(4)$ WZW theory exhibits quasi-long-range correlations. The twisted Rényi correlation of the transfer matrix in Eq.~\ref{eq:twsited1} can therefore be written as 
\begin{widetext}
\begin{align}
&
\frac{
\displaystyle
\int
\mathcal D[\bm m,\bm n]\,
\big[
\bm m(x_0,y_0)\otimes\bm n(x_0,y_0)
\big]
\big[
\bm m(x_0,y_0+N)\otimes\bm n(x_0,y_0+N)
\big]
e^{-\mathcal S[\bm m]
-\mathcal S[\bm n]
-\mathcal S_{\mathrm{int}}}
}{
\displaystyle
\int
\mathcal D[\bm m,\bm n]\,
e^{-\mathcal S[\bm m]
-\mathcal S[\bm n]
-\mathcal S_{\mathrm{int}}}
}
\nonumber\\
&\hspace{2cm}
=
\frac{
\Tr\!\left[
O^\dagger(x)Q^NO(x)Q^N
\right]
}{
\Tr[Q^{2N}]
}
\sim
\frac{1}{|N|^a}.
\label{eq:1dfin}
\end{align}
\end{widetext}
The first line is the correlation of WZW theory, which also renders the type-II strange correlator of the 2d mSPT\cite{you2014wave}. The composite operator
$\bm m(x_0,y_0)\otimes\bm n(x_0,y_0)$ acts simultaneously in the ket and bra sectors. The two operator insertions have the same $x$ coordinate but are separated by $N$ along the $y$ direction. After the spacetime rotation $y\rightarrow t$, this spatial correlation becomes a TRNC of the repeated quantum channel.

\section{Physical Interpretation of TRNC}\label{sec:phys}

We further leverage the TRNC in terms of the time correlation of a quantum channel:
\begin{align}\label{trnc2}
C^{s}(N)
&\equiv 
\frac{\Tr\!\left[\, O^{\dagger}(x)\, Q^{N}\, O(x)\, (Q^{\dagger})^{N}\right]}
{\Tr\!\left[\, Q^{N}(Q^{\dagger})^{N}\right]}
\nonumber\\[6pt]
&=
\frac{\displaystyle \sum_{\sigma}\,
\Tr\!\left[
\Bigl(V(x)\,\mathcal{E}^{N}[\sigma]\,V^{\dagger}(x)\Bigr)\,
\mathcal{E}^{N}\!\Bigl[V^{\dagger}(x)\,\sigma\,V(x)\Bigr]
\right]}
{\displaystyle \sum_{\sigma}\,
\Tr\!\left[\mathcal{E}^{N}[\sigma]\,\mathcal{E}^{N}[\sigma]\right]}
\,, 
\nonumber\\[4pt]
&\text{with}\qquad
O(x)\equiv V(x)\otimes V^{\dagger}(x)\, .
\end{align}
Here we consider the Hermitian-conjugate variant of the TRNC, to be compared with Eq.~\ref{eq:autoco}. This modification does not affect its qualitative behavior in the large-$N$ limit. Here, $\sigma$ denotes the superoperator on the system Hilbert space and need not be a density matrix. For a CPTP map, physical steady states of $\mathcal{E}^{N}$ are positive-semidefinite operators with unit trace, whereas eigenoperators associated with decaying modes are necessarily traceless. Both the stationary and decaying operator sectors must therefore be included in the analysis.

$C^s(N)$ can be viewed as a dynamical probe of the temporal coherence of a charged operator under repeated applications of $\mathcal{E}$. Conceptually, it corresponds to the following protocol: starting from a density matrix $\rho$, we insert a local strong-symmetry charge at position $x$ by conjugation, $\rho\mapsto V^{\dagger}(x)\rho V(x)$. We then evolve under the quantum channel $\mathcal{E}$ for $N$ steps, remove (annihilate) the inserted charge, and compare the resulting process with the depth-$N$ evolution under $\mathcal{E}$ without any charge insertion or annihilation. In this way, $C^s(N)$ measures how much coherence associated with the charged operator survives across two depth-$N$ segments of the evolution, thereby quantifying the resulting operator dynamics and charge correlations.

If the steady state exhibits SWSSB and the $Q$-spectrum is gapped, only the degenerate steady states (one per strong-symmetry sector) contribute, so $C^s(N)$ displays LRO along the replica-$N$ direction. In this case, the operation $V^{\dagger}(x)\rho V(x)$ simply permutes between steady states in different strong-symmetry sectors. If additional gapless modes are present in the $Q$-spectrum, those low-energy modes also contribute and $C^s(N)$ can exhibit power-law tail.
Finally, as noted in Sec.~\ref{sec:LSMchain}, a single application of the quantum channel can already trigger SWSSB, so that the dual 2d mixed state in Eq.~\ref{eq:mSPT} exhibits SWSSB. In this case, the channel’s TRNC (equivalently, the strange correlator of the dual SPT) still exhibits long-range order.

\section{Quantum Channels for Integer-Spin Chains with a Gapped Liouvillian singular spectrum}

In this section, we consider a quantum channel on a spin integer chain that preserves a strong on-site $U(1)$ symmetry (for $S_z$ conservation) and a weak symmetry $T_x \times Z_2$ defined in Eq.~\ref{eq:sym1}.
The onsite Hilbert space consists of two spin-\(\tfrac{1}{2}\) degrees of freedom per unit cell, with basis \(\ket{10}\), \(\ket{01}\), \(\ket{11}\), and \(\ket{00}\). The 1d chain may thus be viewed as having local states that mix spin-1 and spin-0 sectors. Equivalently, this construction can be understood as a spin-\(\tfrac{1}{2}\) chain with explicitly broken translation symmetry, where each unit cell is formed by pairing neighboring even and odd sites. 

\subsection{Gapped Liouvillian singular spectrum} \label{sec:U1gapped}

The integer-spin chain with strong onsite \(U(1)\) symmetry preserving \(S_z\), together with weak \(T_x \times Z_2\) symmetry, is anomaly-free. Our goal is to construct a quantum channel for which the spectrum of \(Q\) is gapless, owing to hydrodynamic modes associated with the strong \(U(1)\) symmetry, while the singular spectrum of \(Q\) is gapped and exhibits a unique dominant eigenstate.

We relabel the onsite Hilbert space as follows:
\begin{align}
|a\rangle &= |00\rangle, \\
|b\rangle &= |\psi_+\rangle = \frac{|01\rangle+|10\rangle}{\sqrt{2}}, \\
|c\rangle &= |\psi_-\rangle = \frac{|01\rangle-|10\rangle}{\sqrt{2}}, \\
|d\rangle &= |11\rangle .
\end{align}
We consider the onsite quantum channel in which $|00\rangle$ and $|11\rangle$ are unchanged, while the symmetric EPR state $|\psi_+\rangle$ decays to the antisymmetric EPR state $|\psi_-\rangle$ with rate $\gamma$.
A convenient Kraus representation is
\begin{align}\label{eq:onsite}
\mathcal Q(\rho)=K_0 \rho K_0^\dagger + K_1 \rho K_1^\dagger ,
\end{align}
with
\begin{align}
K_0
&=
|a\rangle\langle a|
+\sqrt{1-\gamma}\,|b\rangle\langle b|
+|c\rangle\langle c|
+|d\rangle\langle d|, \\
K_1
&=
\sqrt{\gamma}\,|c\rangle\langle b| .
\end{align}
In the basis $(|a\rangle,|b\rangle,|c\rangle,|d\rangle)$, these Kraus operators take the form
\begin{align}
K_0=
\begin{pmatrix}
1&0&0&0\\
0&\sqrt{1-\gamma}&0&0\\
0&0&1&0\\
0&0&0&1
\end{pmatrix},
\qquad
K_1=
\begin{pmatrix}
0&0&0&0\\
0&0&0&0\\
0&\sqrt{\gamma}&0&0\\
0&0&0&0
\end{pmatrix}.
\end{align}

Let Q denote the transfer matrix of the channel in Liouville space. Since the quantum channel acts on-site, we may analyze a single unit cell independently. For each unit cell, the channel has three degenerate steady states,
\begin{align}
\mathrm{span} \{ \ket{a}, \ket{c}, \ket{d} \}.
\end{align}
For a 1d chain of N sites, this implies Q having steady-state degeneracy of $3^N$. However, this extensive degeneracy is an artifact of restricting to a purely on-site channel. Once additional intersite Kraus operators are included, the spectrum of Q becomes gapless within each strong-symmetry sector.

The spectrum of $Q^\dagger Q$ is gapped, with a unique dominant eigenvector:
\begin{align}
\lambda_{\max}
=
(1-\gamma+\gamma^2
+
\gamma\sqrt{(1-\gamma)^2+1})^{N},
\end{align}
which is strictly larger than $1$ for any $\gamma>0$.

The corresponding dominant eigenvector for $Q^\dagger Q$ is:
\begin{align}
\rho_{\max}
=\otimes_i
\frac{
|\psi_+\rangle\langle\psi_+|_i
+
\bigl(1-\gamma+\sqrt{(1-\gamma)^2+1}\bigr)
|\psi_-\rangle\langle\psi_-|_i
}{
2-\gamma+\sqrt{(1-\gamma)^2+1}
}.
\end{align}
Thus, the dominant eigenvector of \(Q^\dagger Q\) is an onsite mixed-state density matrix supported entirely in the \(\{|\psi_+\rangle, |\psi_-\rangle\}\) sector, with \(|\psi_-\rangle\) carrying the larger weight. Importantly, the dominant eigenvector of \(Q^\dagger Q\) is not the same as the steady state of \(Q\).

\subsection{Featureless steady state for the \(S_z=0\) sector}
\label{app:featureless}

We consider the steady state of the quantum channel in Eq.~\ref{eq:onsite} within the \(S_z=0\) sector. This is the only strong-symmetry sector invariant under the weak \(T_x\times Z_2\) symmetry, since the spin-flip operation reverses the total magnetization. In the following, we restrict to the effective onsite Hilbert space
\[
\mathcal H_{\mathrm{eff}}
=
\mathrm{span}\{\ket{a},\ket{c},\ket{d}\},
\]
with the decaying state \(\ket{b}\) removed. The states \(\ket{a}_i\), \(\ket{c}_i\), and \(\ket{d}_i\) carry \(S_z=-1,0,+1\), respectively. The onsite channel alone therefore leaves many configurations in this subspace as steady states.

We now add intersite Kraus operators to lift this degeneracy. On each bond \((i,i+1)\), define the exchange Kraus operators
\[
K_i^{xy}
=
\sqrt{p}\,
|yx\rangle_{i,i+1}\langle xy|,
\qquad
x\neq y,
\qquad
x,y\in\{a,c,d\}.
\]
These operators allow the \(S_z=\pm1\) excitations to move along the chain. We also introduce the fusion Kraus operators
\[
F_i^{ad}
=
\sqrt{q}\,
|cc\rangle_{i,i+1}\langle ad|,
\qquad
F_i^{da}
=
\sqrt{q}\,
|cc\rangle_{i,i+1}\langle da|.
\]
They annihilate a neighboring \(S_z=+1\) and \(S_z=-1\) pair and convert it into two onsite \(S_z=0\) states. Here \(0\leq p,q\) and \(p+q\leq1\).

The remaining Kraus operator is
\[
\begin{aligned}
K_i^0
={}&
\sum_{x=a,c,d}|xx\rangle\langle xx|
+\sqrt{1-p}\sum_{xy\in\{ac,ca,cd,dc\}}|xy\rangle\langle xy|
\\
&+\sqrt{1-p-q}\sum_{xy\in\{ad,da\}}|xy\rangle\langle xy|.
\end{aligned}
\]
The bond channel is
\[
\mathcal E_i(\rho)
=
K_i^0\rho K_i^{0\dagger}
+
\sum_{x\neq y}K_i^{xy}\rho K_i^{xy\dagger}
+
F_i^{ad}\rho F_i^{ad\dagger}
+
F_i^{da}\rho F_i^{da\dagger}.
\]
The global update is obtained by applying these bond channels in a translation-invariant brick-wall circuit with periodic boundary conditions.

This channel preserves the strong \(U(1)\) symmetry generated by
\[
Q=\sum_i S_i^z,
\]
as well as the weak \(T_x\times Z_2\) symmetry. The \(Z_2\) spin flip acts as
\[
X|a\rangle_i=|d\rangle_i,\qquad
X|d\rangle_i=|a\rangle_i,\qquad
X|c\rangle_i=-|c\rangle_i.
\]

For \(p>0\) and \(q>0\), the unique steady state in the total \(S_z=0\) sector is
\[
\rho_{\mathrm{ss}}^{S_z=0}
=
|cc\cdots c\rangle\langle cc\cdots c|.
\]
This is the pure tensor-product state
\(\bigotimes_i|c\rangle_i\), where \(|c\rangle_i\) is an onsite singlet. It is therefore featureless: it preserves the \(U(1)\) symmetry associated with \(S_z\) conservation and the weak \(T_x\times Z_2\) symmetry, and has no long-range order, including SWSSB order. Consequently, it has neither long-range mutual information nor long-range conditional mutual information (CMI)~\cite{zhang2026local,shu2026universal}. This agrees with the statement that an anomaly-free quantum channel can admit a featureless steady state with finite correlation length and finite Markov length in a fixed strong-symmetry sector.

Finally, this example shows that a unique featureless steady state within a fixed strong-symmetry sector does not imply a gapped Liouvillian. Although the steady state has finite correlation length, as diagnosed by the absence of long-range mutual information, and finite Markov length, as diagnosed by the absence of long-range CMI, the Liouvillian spectrum remains gapless in the thermodynamic limit due to strong U(1) symmetry. This differs from the usual equilibrium intuition for closed systems, where a featureless ground state is typically associated with a finite energy gap.

\section{Quantum channel with strong $Z_2$ symmetry}\label{sec:nondeg}
In this section, we consider a 0$d$ quantum channel with strong \(Z_2\) symmetry. Although such channels should exhibit degenerate steady states associated with the even and odd symmetry sectors, the Liouvillian singular spectrum
\(
\Omega = Q^\dagger Q
\)
can nevertheless select a unique dominant eigenvector from one sector.

We study a spin-resolved decay channel for a spinful hardcore boson with local Hilbert space
\begin{align}
\mathcal H
=
\mathrm{span}\Big\{
|0,\uparrow\rangle,\,
|1,\uparrow\rangle,\,
|0,\downarrow\rangle,\,
|1,\downarrow\rangle
\Big\},
\end{align}
where \(0,1\) denote the boson flavor and \(\uparrow,\downarrow\) label the spin.

The quantum channel is
\begin{align}
Q_t(\rho)=K_0\rho K_0^\dagger + K_\uparrow \rho K_\uparrow^\dagger 
+ K_\downarrow \rho K_\downarrow^\dagger,
\end{align}
with Kraus operators
\begin{align}
K_0 &=
|0,\uparrow\rangle\langle 0,\uparrow|
+\sqrt{\eta_\uparrow}\,|1,\uparrow\rangle\langle 1,\uparrow|\nonumber\\
&\quad
+|0,\downarrow\rangle\langle 0,\downarrow|
+\sqrt{\eta_\downarrow}\,|1,\downarrow\rangle\langle 1,\downarrow|,\\
K_\uparrow &= \sqrt{1-\eta_\uparrow}\,|0,\uparrow\rangle\langle 1,\uparrow|,\\
K_\downarrow &= \sqrt{1-\eta_\downarrow}\,|0,\downarrow\rangle\langle 1,\downarrow|.
\end{align}
The decay rate is spin dependent, and we assume \(\eta_\uparrow>\eta_\downarrow\). Notably, this channel preserves \(S_z\), so it has a strong \(Z_2\) symmetry. This is evident from its block-diagonal structure, which separately preserves the spin-up and spin-down sectors.

Each spin sector supports its own steady density matrix,
\begin{align}
\rho_\uparrow^\ast = |0,\uparrow\rangle\langle 0,\uparrow|,
\qquad
\rho_\downarrow^\ast = |0,\downarrow\rangle\langle 0,\downarrow|.
\end{align}
Hence the channel has degenerate steady states as a consequence of the strong \(Z_2\) symmetry. The corresponding eigenvalue is \(1\) in both sectors, independent of the decay rates.

By contrast, the spectrum of \(\Omega = Q^\dagger Q\) is more selective. When \(\eta_\uparrow>\eta_\downarrow\), the largest eigenvalue of \(\Omega\) is nondegenerate and is unique.
Therefore, a strongly \(Z_2\)-symmetric quantum channel can still have a unique dominant eigenvector in the Liouvillian singular spectrum.

Now suppose we further impose both \(S_z\) and \(S_x\) as strong symmetries of the channel. This requires
\begin{align}
\eta_\uparrow=\eta_\downarrow.
\end{align}
Since \(S_z\) and \(S_x\) form a projective representation for a spin-\(\tfrac12\) degree of freedom, this may be viewed as a 0d symmetry anomaly. Under this additional constraint, the leading eigenvalue of \(\Omega\) becomes degenerate again. In this sense, the anomaly enforces a degeneracy in the spectrum of \(\Omega\).

\section{A channel whose steady state breaks weak symmetry}\label{app:weakbreak}

Now we consider a quantum channel for spin 1/2 chain with a strong onsite $U(1)$ symmetry associated with $S_z$ conservation, together with a weak $T_x \times Z_2$ symmetry defined in Eq.~\ref{eq:sym1}. As demonstrated in the main text, the strong $U(1)$ symmetry and the weak $T_x \times Z_2$ symmetry together exhibit a mixed anomaly. Therefore, its steady state cannot be completely featureless: it must exhibit either long-range mutual information, as in a weak-symmetry-breaking steady state, or long-range conditional mutual information, as in an SWSSB steady state. In this appendix, we construct a simple channel whose steady state does not break the strong $U(1)$ symmetry, but instead spontaneously breaks the weak symmetry. 

Consider a spin-$1/2$ chain with onsite basis
\begin{equation}
\mathcal H_i=\mathrm{span}\{\ket{+}_i,\ket{-}_i\},
\end{equation}
where $\ket{+}$ and $\ket{-}$ carry opposite $S_z$ charges. We define a three-site channel acting on each cluster $(i,i+1,i+2)$. The channel leaves the fully polarized and locally antiferromagnetic configurations invariant:
\begin{equation}
+++\to +++,\qquad ---\to ---,
\end{equation}
\begin{equation}
+-+\to +-+,\qquad -+-\to -+-.
\end{equation}
The remaining configurations are rearranged into locally antiferromagnetic patterns:
\begin{equation}
++-\to +-+,\qquad -++\to +-+,
\end{equation}
\begin{equation}
--+\to -+-,\qquad +--\to -+-.
\end{equation}
Thus the channel only changes the order of spins within a three-site cluster. It never changes the total number of $\ket{+}$ and $\ket{-}$ spins, and hence preserves the strong $U(1)$ symmetry.

The Kraus operators that rearrange configurations toward locally antiferromagnetic patterns are
\begin{equation}
J_i^{++-}=\ket{+-+}\bra{++-},
\qquad
J_i^{-++}=\ket{+-+}\bra{-++},
\end{equation}
\begin{equation}
J_i^{--+}=\ket{-+-}\bra{--+},
\qquad
J_i^{+--}=\ket{-+-}\bra{+--}.
\end{equation}
We also include a Kraus operator that keeps the locally stable configurations fixed,
\begin{equation}
K_i^0=P_i^{\mathrm{stable}},
\end{equation}
where
\begin{align}
P_i^{\mathrm{stable}}
=&\ket{+++}\bra{+++}
+\ket{---}\bra{---} \nonumber\\
&+\ket{+-+}\bra{+-+}
+\ket{-+-}\bra{-+-}.
\end{align}
The local quantum channel is therefore
\begin{equation}
\mathcal E_i(\rho)
=
K_i^0\rho K_i^{0\dagger}
+
\sum_{\alpha}J_i^\alpha \rho J_i^{\alpha\dagger},
\end{equation}
with
\begin{equation}
\alpha\in\{++-,-++,--+,+--\}.
\end{equation}

Since each Kraus operator preserves the total $S_z$ charge of the three-site cluster, the channel has a strong $U(1)$ symmetry generated by the total magnetization
\begin{equation}
Q=\sum_i S_i^z .
\end{equation}

The channel also has a weak $Z_2$ spin-flip symmetry. The spin flip acts as
\begin{equation}
X\ket{+}_i=\ket{-}_i,
\qquad
X\ket{-}_i=\ket{+}_i .
\end{equation}
Under this symmetry, the Kraus operators are exchanged in pairs:
\begin{equation}
J_i^{++-}\leftrightarrow J_i^{--+},
\qquad
J_i^{-++}\leftrightarrow J_i^{+--}.
\end{equation}
The projector $P_i^{\mathrm{stable}}$ is also invariant under the spin flip. Therefore the channel as a whole is invariant under the weak $Z_2$ symmetry. Since the same local rule is applied to all translated clusters, the global channel also preserves weak translation symmetry.

Now consider the $S_z=0$ strong-symmetry sector of an even-length chain. In this sector, the steady states are the two Néel states
\begin{equation}
\ket{N_+}=\ket{+-+-\cdots},
\qquad
\ket{N_-}=\ket{-+-+\cdots}.
\end{equation}
These states have fixed total charge $S_z=0$, so they do not break the strong $U(1)$ symmetry. However, such steady state spontaneously breaks the weak $Z_2$ symmetry, giving rise to long-range order and hence long-range mutual information.
 Of course, this construction is fine tuned. Under more generic symmetry-preserving perturbations of the channel, one expects the steady state to more typically develop strong-to-weak symmetry breaking of the strong $U(1)$ symmetry. Nevertheless, the example illustrates the general principle: for a quantum channel with a mixed anomaly between a strong symmetry $S$ and a weak symmetry $G$, the steady state must exhibit long-range CMI or long-range MI, which can appear either as SWSSB of $S$ or as SSB of the weak symmetry $G$.

 \end{document}